\documentclass[11pt,a4paper]{article}
\usepackage[cp1251]{inputenc} 
\usepackage{amsmath,amsthm,amssymb,epsfig,latexsym}
\usepackage{ulem}
\usepackage[numbers,square,sort&compress]{natbib}
\usepackage[english]{babel}
\usepackage{color}

\newtheorem{prop}{Proposition}[section]
\newtheorem{lemma}{Lemma}[section]
\newtheorem{cor}{Corollary}[section]

 \makeatletter
 \@addtoreset{equation}{section}
 \makeatother

\newcommand{\be}[1]{\begin{equation}\label{#1}}
\newcommand{\ba}[1]{\begin{multline}\label{#1}}
\newcommand{\ee}{\end{equation}}
\newcommand{\ea}{\end{eqnarray}}

\def\Izer{{\sf K}}

\def\<{\langle}
\def\>{\rangle}

\def\qed{\hfill$\square$\medskip}

\newcommand{\la}{u}
\newcommand{\muu}{v}
\def\lac{u^{\scriptscriptstyle C}}
\def\lab{u^{\scriptscriptstyle B}}
\def\muc{v^{\scriptscriptstyle C}}
\def\mub{v^{\scriptscriptstyle B}}
\def\blac{\bar u^{\scriptscriptstyle C}}
\def\blab{\bar u^{\scriptscriptstyle B}}
\def\bmuc{\bar v^{\scriptscriptstyle C}}
\def\bmub{\bar v^{\scriptscriptstyle B}}
\newcommand{\bla}{\bar u}
\newcommand{\bmu}{\bar v}

\def\E{{\sf E}}

\def\muu{v}
\newcommand{\so}{{\scriptscriptstyle \rm I}}
\newcommand{\st}{{\scriptscriptstyle \rm I\hspace{-1pt}I}}

\newcommand{\num}{\\\rule{0pt}{20pt}}

\newcommand{\grat}{g^{(0)}}
\newcommand{\frat}{f^{(0)}}
\newcommand{\hrat}{h^{(0)}}
\newcommand{\trat}{t^{(0)}}
\newcommand{\Izerl}{\Izer^{(l)}}
\newcommand{\Izerr}{\Izer^{(r)}}
\newcommand{\Izerlr}{\Izer^{(l,r)}}
\newcommand{\Izerrl}{\Izer^{(r,l)}}
\newcommand{\RHCl}{{\sf Z}^{(l)}}
\newcommand{\RHCr}{{\sf Z}^{(r)}}
\newcommand{\RHClr}{{\sf Z}^{(l,r)}}

\newcommand{\tr}{\mathop{\rm tr}}
\newcommand{\diag}{\mathop{\rm diag}}

\begin{document}

\vspace{20pt}

\begin{center}
\begin{LARGE}
 {\bf Scalar products in  $GL(3)$-based models with  \\[1.2ex]
 trigonometric $R$-matrix.  Determinant representation\\[1.2ex]
 }
\end{LARGE}
\vspace{50pt}
\begin{large}
{N.~A.~Slavnov\footnote{nslavnov@mi.ras.ru}}
\end{large}


\vspace{4mm}

{\it Steklov Mathematical Institute,
Moscow, Russia}

\end{center}


\vspace{4mm}


\begin{abstract}
We study quantum integrable $GL(3)$-based models with a  trigonometric $R$-matrix  solvable by the nested algebraic
Bethe ansatz. We derive a determinant representation for a special case of scalar products of Bethe vectors. This representation
allows one to find a determinant formula for the form factor of one of the monodromy matrix entries. We also point out essential
difference between form factors in the models with the trigonometric $R$-matrix  and their analogs in $GL(3)$-invariant models.
\end{abstract}

\vspace{1cm}

\vspace{2mm}

\section{Introduction}

The algebraic Bethe ansatz \cite{FadST79,FadLH96,KulS79,FadT79} allows one to obtain the spectra of quantum Hamiltonians of many models of physical interest. The calculation of correlation functions and form factors also can be performed in the framework
of this method. The last problem in many cases can be reduced to the calculation of scalar products of Bethe
vectors.  A systematic study of scalar products in  quantum integrable $GL(3)$-based  models with
a trigonometric $R$-matrix \cite{KulS80,PerS81} was initiated in  \cite{PakRS13c,PakRS14b}.

Recently, the form factors of the monodromy matrix entries in  models with $GL(3)$-invariant
$R$-matrix were studied in works \cite{BelPRS12b,BelPRS13a,PakRS14d,PakRS14e,PakRS15a}. These form factors can be used for  calculating correlation functions in a wide class of Bethe ansatz solvable models.
It was shown in \cite{PakRS15a} that all form factors of the monodromy matrix entries are related to each other. Actually, it is enough
to calculate only one of them. All the others can be obtained from this initial one within the special limits of Bethe parameters. This method was
developed in \cite{PakRS15a}, where it was called {\it a zero modes method}. In fact, this method allows one to obtain all the  form factors
from a scalar product of Bethe vectors of a special type.

A question arises about a generalization of these results to the models with trigonometric (or $q$-deformed) $R$-matrix.
The trigonometric quantum $R$-matrix for the $GL(3)$-based models  has the following form:
\begin{equation}\label{UqglN-R}
\begin{split}
R(u,v)\ =\ f(u,v)&\ \sum_{1\leq i\leq 3}\E_{ii}\otimes \E_{ii}\ +\
\sum_{1\leq i<j\leq 3}(\E_{ii}\otimes \E_{jj}+\E_{jj}\otimes \E_{ii})
\\
+\ &\sum_{1\leq i<j\leq 3}
(u\,g(u,v) \E_{ij}\otimes \E_{ji}+ v\,g(u,v)\E_{ji}\otimes \E_{ij}),
\end{split}
\end{equation}
where the rational functions $f(u,v)$ and $g(u,v)$   are
\begin{equation}\label{fgg}
f(u,v)=\frac{qu-q^{-1}v}{u-v},\quad g(u,v)=\frac{q-q^{-1}}{u-v}\,,
\end{equation}
and $q$ is a complex number (a deformation parameter). The matrices $(\E_{ij})_{lk}=\delta_{il}\delta_{jk}$, $i,j,l,k=1,2,3$  of the size $3\times3$
have a unit in the intersection of $i$th row and $j$th column and zero matrix elements elsewhere.

For the $GL(2)$-based models $q$-deformation of the $R$-matrix does not lead to serious problems in calculating the scalar
products of Bethe vectors. In particular, there exists a universal determinant representation for the scalar product of
an arbitrary Bethe vector with an eigenstate of the transfer matrix \cite{Sla89}. Being written in terms of the functions \eqref{fgg}
this representation is valid both for the models with $GL(2)$-invariant $R$-matrix and for the  models with the $q$-deformed symmetry.
In the first case one should only take the scaling limit in \eqref{fgg}: $u=1+\epsilon u'$, $v=1+\epsilon v'$, $q=1+\epsilon c/2$,
$\epsilon \to 0$.

One could expect that this similarity is preserved in the
models described by the $q$-defor\-ma\-tion of higher rank algebras. This expectation was partly confirmed in
the recent works devoted to the analysis of scalar products in the $GL(3)$-based models with the trigonometric $R$-matrix  \cite{PakRS13c,PakRS14b}.
In particular, there was  obtained a $q$-deformed generalization of the {\it sum formula} \cite{Res86} for the scalar product
of generic Bethe vectors. In this representation the scalar products are given as sums over partitions of the sets of Bethe parameters. It is
known, however, that this  type of representations are not convenient either for analytical calculations of correlation functions or
for their numeric analysis.  It would be desirable to reduce the sums over partitions of Bethe parameters to
determinant formulas similar to the ones obtained in \cite{BelPRS12b,BelPRS13a,PakRS14d,PakRS14e,PakRS15a} for the models with $GL(3)$-invariant
$R$-matrix. The main goal of this paper is to obtain these  representations.

To our great surprise it turned out that the $q$-deformation of $GL(3)$-invariance leads to significant changes at the level of determinant representations for form factors. In the present paper we obtain this representation for  form factor of only one entry of the monodromy matrix. Furthermore, we show
that obtaining determinant representations for other form factors is impossible, at least in the framework of the scheme considered below. Instead, one
can derive determinant formulas for so called {\it twisted form factors}, where one of the vectors is an eigenstate of a twisted transfer matrix (see section~\ref{S-TTM}) with a special twist parameter.

The paper is organized as follows. In section~\ref{S-N} we describe the model under consideration and introduce
a necessary notation. Section~\ref{S-BV} is devoted to the description of Bethe vectors and their scalar products. In  section~\ref{S-TTM} we
introduce a notion of a twisted transfer matrix that we use for  calculating form factors.
In  section~\ref{S-MR} we present the main result of the paper: a determinant representation for a special type of the scalar
product of Bethe vectors in the $GL(3)$-based models with the trigonometric $R$-matrix. The remaining part of the paper is devoted to the proof of these formulas. In section~\ref{S-SI} we describe some properties of the Izergin determinant and formulate several lemmas used in the next sections. In sections~\ref{S-DDD}
and~\ref{S-FS} we derive a determinant formula for the scalar product of twisted on-shell and usual on-shell Bethe vectors. Section~\ref{S-FinDis}
is devoted to the conclusive discussions.  Appendices~\ref{A-Long-Det}~and~\ref{A-Wau}
contain the proofs of the lemmas formulated in section~\ref{S-SI}.

\section{Notation\label{S-N}}

In this paper we consider  quantum integrable models with a $3\times 3$ monodromy matrix
$T(u)$. It satisfies standard commutation relation ($RTT$-relation):
\begin{equation}\label{RTT}
R(u,v)\cdot (T(u)\otimes \mathbf{1})\cdot (\mathbf{1}\otimes T(v))=
(\mathbf{1}\otimes T(v))\cdot (T(u)\otimes \mathbf{1})\cdot R(u,v)\,.
\end{equation}
The entries $T_{ij}(u)$ of the monodromy matrix act in a quantum space $V$ and form the quadratic algebra with commutation
relations given by \eqref{RTT}.
We assume that the vector space $V$
 possesses a highest weight vector $|0\rangle\in V$ such that:
\begin{equation}\label{rsba}
T_{ij}(u)|0\rangle =0,\quad i>j,\quad T_{ii}(u)|0\rangle= \lambda_i(u)|0\rangle\,,\qquad \lambda_i(u)\in\mathbb{C}[[u,u^{-1}]]\,.
\end{equation}
We also assume that the operators $T_{ij}(u)$ act in a dual space $V^*$ with
a vector $\langle 0|\in V^*$ such that:
\begin{equation}\label{drsba}
\langle 0|T_{ij}(u) =0,\quad i<j,\quad  \langle 0|T_{ii}(u)= \lambda_i(u)\langle 0|\,,
\end{equation}
and  $\lambda_i(u)$ are the same as in \eqref{rsba}. Actually, it is always possible to normalize
the monodromy matrix $T(u)\to \lambda_2^{-1}(u)T(u)$ so as to deal only with the ratios
 \be{ratios}
 r_1(u)=\frac{\lambda_1(u)}{\lambda_2(u)}, \qquad  r_3(u)=\frac{\lambda_3(u)}{\lambda_2(u)}.
 \ee
Below we assume that $\lambda_2(u)=1$.

Apart from the functions $g(x, y)$ and  $f(x, y)$ \eqref{fgg} we also use two other auxiliary functions
\be{desand}
h(x,y)=\frac{f(x,y)}{g(x,y)}=\frac{qx-q^{-1}y}{q-q^{-1}},\qquad  t(x,y)=\frac{g(x,y)}{h(x,y)}=\frac{(q-q^{-1})^2}{(x-y)(qx-q^{-1}y)}.
\ee
The following obvious properties of the functions introduced above are useful:
\be{formulas}
\begin{aligned}
&h(xq^{-2},y)=q^{-1}g^{-1}(x,y),  \qquad &g(x,yq^{-2})=q h^{-1}(x,y),\qquad &g(xq^{-2},x)=-qx^{-1},\\
&t(xq^{-2},y)=q^2t(y,x),\qquad &t(x,yq^{2})=q^{-2}t(y,x),\qquad &f(xq^{-2},y)=f^{-1}(y,x).
\end{aligned}
\ee

Below we permanently deal with sets of variables and their partitions into subsets. We denote sets of variables by bar: $\bla$, $\bmu$ and so on,
\begin{equation}\label{set111}
\bla = \left\{u_1,\ldots,u_a\right\},\quad \bmu=\left\{v_1,\ldots,v_b\right\}\,.
\end{equation}
If necessary, the cardinalities of the sets are described in special comments
after the formulas. The individual elements
of the sets are denoted by subscripts: $u_j$, $v_k$ etc. Special notation
$\bar u_j$, $\bar v_k$ and similar ones are used for subsets with one element omitted: $\bar u_j=\bar u\setminus u_j$,
$\bar v_k=\bar v\setminus v_k$ and so on.

If a set of variables is
multiplied by a number $\alpha\bla$ (in particular, $\bla q^{\pm 2}$), then it means that all the
elements of the set are multiplied by this number:
\begin{equation}\label{set-numb}
\alpha\bla = \left\{\alpha u_1,\ldots,\alpha u_a\right\},\qquad
\bmu q^{\pm 2}=\left\{v_1q^{\pm 2},\ldots,v_bq^{\pm 2}\right\}\,.
\end{equation}
If the order of the elements in a set  is essential, then we assume
that the elements  are ordered in such a way that the sequence of
their subscripts is strictly increasing. We call this ordering  a natural order.

A union of sets is denoted by braces, for example, $\{\bar w,\bla\}=\bar\eta$. Partitions of sets into disjoint subsets are denoted by
the symbol $\Rightarrow$, and the subsets are numerated by roman numbers. For example, notation $\bar u \Rightarrow \{\bar u_{\so}, \bar u_{\st}\}$ means that the set $\bla$ is divided into two subsets $\bar u_{\so}$ and $\bar u_{\st}$, such that
$\bar u_{\so}\cap \bar u_{\st}=\emptyset$ and $\{\bar u_{\so}, \bar u_{\st}\}=\bla$. Similarly, notation $\bar\eta\Rightarrow
\{\bar\eta_{\rm i},\bar\eta_{\rm ii},\bar\eta_{\rm iii}\}$ means that the set $\bar\eta$ is divided into three subsets
with pair-wise empty intersections and $\{\bar\eta_{\rm i},\bar\eta_{\rm ii},\bar\eta_{\rm iii}\}=\bar\eta$.

Just like in \cite{PakRS13c} we use a shorthand notation for  products with respect to sets of
variables:
 \begin{equation}\label{SH-prod}
 T_{ij}(\bar w)=\prod_{w_k\in\bar w}   T_{ij}(w_k);\quad
  r_1(\bar v_k)= \prod_{\substack{v_j\in\bar v\\v_j\ne v_k}} r_1(v_j);\quad
 r_3(\bar v_{\so})= \prod_{v_j\in\bar v_{\so}} r_3(v_j).
 \end{equation}
That is, if the operator $T_{ij}$ or the functions $r_k$ depend on a set of variables, this means that one should
take the product of the operators or the scalar functions with respect to the corresponding set. The same convention will be used for the products of functions $f(u,v)$, $h(u,v)$, and $t(u,v)$:
\begin{equation}\label{SH-prod1}
 h(u, \bar w)= \prod_{w_j\in\bar w} h(u, w_j);\quad t(\bar v_k,u)= \prod_{\substack{v_j\in\bar v\\v_j\ne v_k}} t(v_j,u);\quad
 f(\bar u_{\so},\bar v_{\st})=\prod_{u_j\in\bar u_{\so}}\prod_{v_k\in\bar v_{\st}} f(u_j,v_k).
 \end{equation}
In addition to the standard shorthand notations we will also use
\be{def-pp}
\mathfrak{p}(\bar x)=\prod_{x_k\in\bar x} x_k\,.
\ee

\section{Bethe vectors \label{S-BV}}

Bethe vectors are special polynomials in the operators $T_{jk}(w)$ with $j<k$ applied to the vector $|0\rangle$.
Similarly, dual Bethe vectors are special polynomials in the operators $T_{jk}(w)$ with $j>k$ applied to $\langle0|$.
Their explicit form in the $GL(N)$-based models with the trigonometric $R$-matrix  was found in \cite{PakRS14ab}. We denote Bethe vectors and their dual
ones respectively  by $\mathbb{B}^{a,b}(\bla;\bmu)$ and $\mathbb{C}^{a,b}(\bla;\bmu)$, stressing that they depend on two sets of variables
$\bla$ and $\bmu$. These variables are called Bethe parameters. The superscripts $a$ and $b$ show the cardinalities
of the sets $\bla$ and $\bmu$: $\#\bla=a$, $\#\bmu=b$.

A Bethe vector\footnote{For simplicity below we do not distinguish between Bethe vectors and dual ones, if
it does not cause a misunderstanding.} becomes an eigenvector of the transfer matrix $\tr T(u)$ if the Bethe
parameters satisfy a system of Bethe equations
\be{AEigenS-1}
r_1(\la_{j})=\frac{f(\la_{j},\bla_{j})}{f(\bla_{j},\la_{j})}f(\bmu,\la_{j}),\qquad
r_3(\muu_{j})=\frac{f(\bmu_{j},\muu_{j})}{f(\muu_{j},\bmu_{j})}f(\muu_{j},\bla),
\ee
and we recall that $\bla_{j}=\bla\setminus u_j$, $\bmu_{j}=\bmu\setminus v_j$. In this case we call the corresponding vector
{\it on-shell} Bethe vector. Observe that if $\bla$ and $\bmu$ satisfy \eqref{AEigenS-1}, then for arbitrary partitions
$\bla\Rightarrow\{\bla_{\so},\bla_{\st}\}$ and $\bmu\Rightarrow\{\bmu_{\so},\bmu_{\st}\}$
\be{AEigenS-1p}
r_1(\bla_{\so})=\frac{f(\bla_{\so},\bla_{\st})}{f(\bla_{\st},\bla_{\so})}f(\bmu,\bla_{\so}),\qquad
r_3(\bmu_{\so})=\frac{f(\bmu_{\st},\bmu_{\so})}{f(\bmu_{\so},\bmu_{\st})}f(\bmu_{\so},\bla).
\ee
To obtain \eqref{AEigenS-1p} it is enough to take the products in \eqref{AEigenS-1} with respect to the subsets
$\bla_{\so}$ and $\bmu_{\so}$.

If the system of Bethe equations \eqref{AEigenS-1} holds, then
\be{Left-act}
\tr T(w)\mathbb{B}^{a,b}(\bla;\bmu) = \tau(w|\bla,\bmu)\,\mathbb{B}^{a,b}(\bla;\bmu),\qquad
\mathbb{C}^{a,b}(\bla;\bmu)\tr T(w) = \tau(w|\bla,\bmu)\,\mathbb{C}^{a,b}(\bla;\bmu),
\ee
where
\be{tau-def}
\tau(w|\bla,\bmu)=r_1(w)f(\bla,w)+f(w,\bla)f(\bmu,w)+r_3(w)f(w,\bmu).
\ee

Scalar products of Bethe vectors are defined as follows:
 \be{SP-def-gen}
 S_{a,b}\equiv S_{a,b}(\blac,\bmuc;\blab,\bmub)=
 \mathbb{C}^{a,b}(\blac;\bmuc)\mathbb{B}^{a,b}(\blab;\bmub).
 \ee
Here the vectors  $\mathbb{C}^{a,b}(\blac;\bmuc)$ and $\mathbb{B}^{a,b}(\blab;\bmub)$ supposed to be generic Bethe vectors,
i.e. their Bethe parameters are generic complex numbers. We have added the superscripts $C$ and $B$
to the sets $\bar u$, $\bar v$ in order to stress that the vectors
$\mathbb{C}^{a,b}(\blac;\bmuc)$ and $\mathbb{B}^{a,b}(\blab;\bmub)$ may depend on different sets of parameters. In other words, unless explicitly
specified, the variables $\{\blab, \bmub\}$ in $\mathbb{B}^{a,b}$ and
$\{\blac, \bmuc\}$ in $\mathbb{C}^{a,b}$ are not supposed to be equal.

The main goal of this paper is to find determinant representations for form factors of the diagonal entries $T_{ii}(z)$ of the monodromy matrix.
We denote them by  $\mathcal{F}_{a,b}^{(i,i)}(z)$ and define as
 \be{SP-deFF-gen}
 \mathcal{F}_{a,b}^{(i,i)}(z)\equiv\mathcal{F}_{a,b}^{(i,i)}(z|\blac,\bmuc;\blab,\bmub)=
 \mathbb{C}^{a,b}(\blac;\bmuc)T_{ii}(z)\mathbb{B}^{a,b}(\blab;\bmub),
 \ee
where both $\mathbb{C}^{a,b}(\blac;\bmuc)$ and $\mathbb{B}^{a,b}(\blab;\bmub)$ are on-shell
Bethe vectors. The parameter $z$ is an arbitrary complex  number.

\section{Twisted transfer matrix\label{S-TTM}}

One can easily relate the form factors of the diagonal matrix elements $T_{ii}(z)$ with scalar products of a special type
\cite{Kor82,IzeK84,KitMST05,BelPRS12b,BelPRS13a}. For this we introduce the  notion of a twisted monodromy matrix and a twisted transfer matrix.

Let $\bar\kappa$ be a set of three complex numbers (twist parameters) $\bar\kappa=\{\kappa_1,\kappa_2,\kappa_3\}$.
We define the  twisted monodromy matrix as $T_{\bar\kappa}=\hat\kappa T$, where
$\hat\kappa=\diag(\kappa_1,\kappa_2,\kappa_3)$.
 It is easy to check that the tensor square of $\hat\kappa$ commutes with the $R$-matrix:
$[\hat\kappa_1\hat\kappa_2,R_{12}(u,v)]=0$, and therefore the twisted monodromy matrix satisfies the algebra \eqref{RTT}
\begin{equation}\label{tRTT}
R(u,v)\cdot (T_{\bar\kappa}(u)\otimes \mathbf{1})\cdot (\mathbf{1}\otimes T_{\bar\kappa}(v))=
(\mathbf{1}\otimes T_{\bar\kappa}(v))\cdot (T_{\bar\kappa}(u)\otimes \mathbf{1})\cdot R(u,v)\,.
\end{equation}
Thus, the
eigenvectors of the twisted transfer matrix $\tr  T_{\bar\kappa}(w)$ can be found within the framework of the standard scheme.
We call them twisted (dual) on-shell Bethe vectors. Like the standard on-shell vectors, they can be parameterized by sets of complex parameters
satisfying twisted Bethe equations
\be{ATEigenS-1}
r_1(\la_{j})=\frac{\kappa_2}{\kappa_1}\frac{f(\la_{j},\bla_{j})}{f(\bla_{j},\la_{j})}f(\bmu,\la_{j}),
\qquad r_3(v_{j})=\frac{\kappa_2}{\kappa_3}\frac{f(\bmu_{j},v_{j})}{f(v_{j},\bmu_{j})}f(v_{j},\bla).
\ee
Similarly to \eqref{AEigenS-1p} one can find
for arbitrary partitions
$\bla\Rightarrow\{\bla_{\so},\bla_{\st}\}$ and $\bmu\Rightarrow\{\bmu_{\so},\bmu_{\st}\}$
\be{ATEigenS-1p}
r_1(\bla_{\so})=\left(\frac{\kappa_2}{\kappa_1}\right)^{k_{\so}}\frac{f(\bla_{\so},\bla_{\st})}{f(\bla_{\st},\bla_{\so})}f(\bmu,\bla_{\so}),\qquad
r_3(\bmu_{\so})=\left(\frac{\kappa_2}{\kappa_3}\right)^{n_{\so}}\frac{f(\bmu_{\st},\bmu_{\so})}{f(\bmu_{\so},\bmu_{\st})}f(\bmu_{\so},\bla),
\ee
where $k_{\so}=\#\bla_{\so}$ and $n_{\so}=\#\bmu_{\so}$.

In this paper we will also consider twisted from factors. They are still defined by \eqref{SP-deFF-gen}, but one of vectors
(for instance, $\mathbb{C}^{a,b}(\blac;\bmuc)$) is a twisted on-shell Bethe vector. Calculating the  twisted and usual form factors of the matrix elements
$T_{ii}(z)$  can be reduced to the scalar products via the following method.

Let $\tr T_{\bar\kappa}(z)$ be the twisted transfer matrix and $\tr T(z)$ be the standard transfer matrix.
Consider
\be{Qm}
Q_{\bar\kappa}(z)=\mathbb{C}^{a,b}(\blac;\bmuc) \bigl(\tr T_{\bar\kappa}(z)-\tr T(z)\bigr)\mathbb{B}^{a,b}(\blab;\bmub),
\ee
where $\mathbb{C}^{a,b}(\blac;\bmuc)$ and $\mathbb{B}^{a,b}(\blab;\bmub)$ are twisted and standard on-shell
vectors, respectively.  Obviously
\be{Qm-0}
Q_{\bar\kappa}(z)=\mathbb{C}^{a,b}(\blac;\bmuc)\sum_{j=1}^3(\kappa_j-1)T_{jj}(z) \mathbb{B}^{a,b}(\blab;\bmub).
\ee
Let $i$ be a fixed number from the set $\{1,2,3\}$, and let $\bar\kappa_i=1$ (that is $\kappa_j=1$ for $j\ne i$). Then we obtain
\be{Qm-FF}
\mathcal{F}^{(i,i)}_{a,b;\kappa_i}(z|\blac,\bmuc;\blab,\bmub)=\frac{Q_{\bar\kappa}(z)}{\kappa_i-1},
\ee
where $\mathcal{F}^{(i,i)}_{a,b;\kappa_i}(z)$ denotes a twisted form factor corresponding to the twist parameters $\bar\kappa_i=1$.

On the other hand
\be{Qm-1}
Q_{\bar\kappa}(z)=\bigl(\tau_{\bar\kappa}(z|\blac;\bmuc)-\tau(z|\blab;\bmub)\bigr)\;\mathbb{C}^{a,b}(\blac;\bmuc)\mathbb{B}^{a,b}(\blab;\bmub),
\ee
where $\tau_{\bar\kappa}(z|\blac;\bmuc)$ is the eigenvalue of the
twisted transfer matrix on the  vector $\mathbb{C}^{a,b}(\blac;\bmuc)$:
\be{ttau-def}
\tau_{\bar\kappa}(z|\blac,\bmuc)=\kappa_1r_1(z)f(\blac,z)+\kappa_2 f(z,\blac)f(\bmuc,z)+\kappa_3r_3(z)f(z,\bmuc).
\ee
Thus, we arrive at
\be{Tw-FF}
\mathcal{F}^{(i,i)}_{a,b;\kappa_i}(z|\blac,\bmuc;\blab,\bmub)=\frac{\tau_{\bar\kappa}(z|\blac;\bmuc)-\tau(z|\blab;\bmub)}{\kappa_i-1}
\,\mathbb{C}^{a,b}(\blac;\bmuc)\mathbb{B}^{a,b}(\blab;\bmub).
\ee

In the limit $\kappa_i\to 1$ we obtain the usual form factor of $T_{ii}(z)$   \cite{KitMST05,BelPRS12b,BelPRS13a}
\be{Qm-4}
\mathcal{F}_{a,b}^{(i,i)}(z|\blac,\bmuc;\blab,\bmub)=\frac{d}{d\kappa_i}\,\bigl(\tau_{\bar\kappa}(z|\blac;\bmuc)-\tau(z|\blab;\bmub)\bigr)\;
\mathbb{C}^{a,b}(\blac;\bmuc)\mathbb{B}^{a,b}(\blab;\bmub)\Bigl.\Bigr|_{\bar\kappa=1}.
\ee
Here the symbol $\bar\kappa=1$ means that $\kappa_j=1$, $j=1,2,3$. Thus,  evaluating twisted and usual form factors
is reduced to the computation of the scalar product between the twisted and usual on-shell Bethe vectors.

\section{Main results\label{S-MR}}

We obtain determinant representations for the scalar product of the twisted and usual on-shell Bethe vectors in two cases: $\kappa_3/\kappa_1=1$ and $\kappa_3/\kappa_1=q^2$. In both cases $\kappa_2$ remains an arbitrary complex number. We denote the corresponding scalar products by $S^{(1)}_{a,b}$
and $S^{(q^2)}_{a,b}$, respectively. In order to describe  determinant representations we first introduce for an arbitrary set of variables
$\bar w=\{w_1,\dots,w_n\}$
\be{def-Del}
\Delta_n(\bar w)
=\prod_{j>k}^n g(w_j,w_k),\qquad \Delta'_n(\bar w)=\prod_{j<k}^n g(w_j,w_k).
\ee
Let also
\be{Ch}
C_h=h(\bmuc,\bmuc)h(\bmuc,\blab)h(\blab,\blab).
\ee
Then we introduce an $a\times(a+b)$ matrix ${\sf N}^{(u)}(\lac_j,x_k)$
\be{Nu-def}
{\sf N}^{(u)}(\lac_j,x_k)=
(-1)^{a-1} t(\lac_j,x_k) \frac{r_1(x_k)}{f(\bmuc,x_k)}\frac{h(\blac,x_k)}{h(x_k,\blab)}
+\frac{\kappa_2}{\kappa_1}t(x_k,\lac_j)\frac{h(x_k,\blac)}{h(x_k,\blab)},
\ee
and a $b\times(a+b)$ matrix ${\sf N}^{(v)}(\mub_j,x_k)$
\be{Nv-def}
{\sf N}^{(v)}(\mub_j,x_k)= (-1)^{b-1} t(x_k,\mub_j)\frac{r_3(x_k)}{f(x_k,\blab)}\frac{h(x_k,\bmub)}
{h(\bmuc,x_k)}
+t(\mub_j,x_k)\frac{h(\bmub,x_k)}{h(\bmuc,x_k)}.
\ee
In these formulas
\be{x-def}
\bar x=\{\blab,\bmuc\}=\{\lab_1,\dots\lab_a,\muc_1,\dots,\muc_b\}.
\ee

\begin{prop}\label{Propos-1}
The determinant representation for $S^{(1)}_{a,b}$ has the following form:
\be{FF1-ans-m1}
S^{(1)}_{a,b}=  \mathfrak{p}(\bmub)\mathfrak{p}(\blab) C_h\Delta'_{a}(\blac)
 \Delta'_b(\bmub)\Delta_{a+b}(\bar x)\;\det_{a+b}{\sf N},
\ee
where
\be{mat-N}
\begin{array}{ll}
{\sf N}_{j,k}={\sf N}^{(u)}(\lac_j,x_k),&j=1,\dots,a,\\
{\sf N}_{j+a,k}={\sf N}^{(v)}(\mub_j,x_k),&j=1,\dots,b,
\end{array}
\qquad \{x_1,\dots,x_{a+b}\}=\{\lab_1,\dots\lab_a,\muc_1,\dots,\muc_b\}.
\ee
\end{prop}

Comparing this result with the analogous scalar product in the models with $GL(3)$-invariant $R$-matrix \cite{BelPRS12b} we see that up to the
trivial prefactor $\mathfrak{p}(\bmub)\mathfrak{p}(\blab)$ it has exactly the same form. One simply should replace the rational functions
$g(u,v)$, $f(u,v)$, $h(u,v)$, and $t(u,v)$ by their analogs in $GL(3)$-invariant models:
\be{rat-analog}
\begin{aligned}
g(u,v)&\to \grat(u,v)=\frac{c}{u-v},\qquad &f(u,v)&\to \frat(u,v)=\frac{u-v+c}{u-v},\\
h(u,v)&\to \hrat(u,v)=\frac{u-v+c}{c},\qquad &t(u,v)&\to \trat(u,v)=\frac{c^2}{(u-v)(u-v+c)},
\end{aligned}
\ee
where $c$ is a constant. This correspondence is quite similar to the one that we have for the $GL(2)$ case.

Similarly to \eqref{mat-N} we define a matrix $\widetilde{\sf N}$ as
\be{mat-tN}
\begin{array}{ll}
\widetilde{\sf N}_{j,k}=x_k{\sf N}^{(u)}(\lac_j,x_k),&j=1,\dots,a,\\
\widetilde{\sf N}_{j+a,k}={\sf N}^{(v)}(\mub_j,x_k),&j=1,\dots,b.
\end{array}
\ee

\begin{prop}\label{Propos-2}
The determinant representation for $S^{(q^2)}_{a,b}$ has the following form:
\be{FF1-ans-mq}
S^{(q^2)}_{a,b}=  \mathfrak{p}(\bmub) C_h\Delta'_{a}(\blac)
 \Delta'_b(\bmub)\Delta_{a+b}(\bar x)\;\det_{a+b}\widetilde{\sf N}.
\ee
\end{prop}

It is clear that in the scaling limit $q\to 1$ representations \eqref{FF1-ans-mq} and \eqref{FF1-ans-m1} coincide and give the determinant
formula for the scalar product of the twisted and usual on-shell vectors in the models with $GL(3)$-invariant $R$-matrix \cite{BelPRS12b}.

{\sl Remark.\label{Rem-ii}}  The matrix elements ${\sf N}^{(u)}(\lac_j,x_k)$ \eqref{Nu-def} and ${\sf N}^{(v)}(\mub_j,x_k)$ \eqref{Nv-def}
require additional definition if $\blac\cap\blab\ne\emptyset$ or $\bmuc\cap\bmub\ne\emptyset$. Let for definiteness
$\lac_a=\lab_a$. The matrix element ${\sf N}^{(u)}(\lac_a,\lab_a)$ depends on the functions $t(\lac_a,\lab_a)$ and
$t(\lab_a,\lac_a)$, which have a pole at $\lab_a=\lac_a$. However one can easily check that the residue in this pole vanishes
due to the twisted Bethe equations \eqref{ATEigenS-1}, and hence, the limit of \eqref{Nu-def} is finite:
\begin{multline}\label{diag-el1}
{\sf N}^{(u)}(\lac_a,\lab_a)\Bigr|_{\lac_a=\lab_a}=\frac{\kappa_2}{\kappa_1\lac_a}\frac{h(\lac_a,\blac)}{h(\blac_a,\blab)}\Biggl[
(q^{-1}-q)\frac{r'_1(\lac_a)}{r_1(\lac_a)}\\
+\sum_{\ell=1}^{a-1}\frac{(q+q^{-1})\lac_\ell}{h(\lac_a,\lac_\ell)h(\lac_\ell,\lac_a)}+\sum_{i=1}^b\muc_i t(\muc_i,\lac_a)\Biggr],
\end{multline}
where $r'_1(\lac_a)$ is the derivative of $r_1(\lac_a)$. It is this expression for ${\sf N}^{(u)}(\lac_a,\lab_a)$
that needs to be used in the case $\lac_a=\lab_a$. Similarly, if
$\muc_{b}=\mub_b$, then
\begin{multline}\label{diag-el2}
{\sf N}^{(v)}(\mub_b,\muc_b)\Bigr|_{\muc_b=\mub_b}=\frac{1}{\mub_b}\frac{h(\bmub,\mub_b)}{h(\bmuc,\mub_b)}\Biggl[
(q-q^{-1})\frac{r'_3(\mub_b)}{r_3(\mub_b)}\\
-\sum_{i=1}^{b-1}\frac{(q+q^{-1})\mub_i}{h(\mub_b,\mub_i)h(\mub_i,\mub_b)}+\sum_{\ell=1}^a\lab_\ell t(\mub_b,\lab_\ell)\Biggr],
\end{multline}
where $r'_3(\mub_b)$ is the derivative of $r_3(\mub_b)$.

Note that in the  particular case $\blac=\blab$, $\bmuc=\bmub$, and $\bar\kappa=1$ representation \eqref{FF1-ans-m1} for $S^{(1)}_{a,b}$ gives
the square of the norm of the on-shell Bethe vector. In this case one should use \eqref{diag-el1} and \eqref{diag-el2} for the diagonal entries
of the matrix ${\sf N}$.

We would like to stress once more that in the obtained results the parameter $\kappa_2$ is an arbitrary complex number, while the ratio
of the parameters $\kappa_1$ and $\kappa_3$ is fixed. Since the final result depends  on the ratios of $\kappa_i$ only, one can say
that $\kappa_1$ and $\kappa_3$ are fixed. This means that using \eqref{Qm-4} we can obtain a determinant representation for the form factor of the
operator $T_{22}(z)$
\be{F22-res}
\mathcal{F}_{a,b}^{(2,2)}(z|\blac,\bmuc;\blab,\bmub)=\frac{d}{d\kappa_2}\,\bigl(\tau_{\bar\kappa}(z|\blac;\bmuc)-\tau(z|\blab;\bmub)\bigr)\;
S^{(1)}_{a,b}(\blac,\bmuc;\blab,\bmub)\Bigl.\Bigr|_{\kappa_2=1}.
\ee
However, we cannot obtain determinant representations for the form factors of the
operators $T_{11}(z)$ and $T_{33}(z)$. For these operators we know only the twisted form factors with a fixed values of the twist parameters, for example
\be{Tw-FF1}
\mathcal{F}^{(3,3)}_{a,b;q^2}(z|\blac,\bmuc;\blab,\bmub)=\frac{\tau_{\bar\kappa}(z|\blac;\bmuc)-\tau(z|\blab;\bmub)}{q^2-1}
\,S^{(q^2)}_{a,b}(\blac,\bmuc;\blab,\bmub),
\ee
where $\bar\kappa=\{1,1,q^2\}$.

The remainder of this work is devoted to the proof of propositions~\ref{Propos-1} and~\ref{Propos-2}.

\section{Summation identities\label{S-SI}}

The central object of the theory of scalar products in the $GL(2)$-based models is the  Izergin determinant $\Izer_k(\bar x|\bar y)$
\cite{Ize87}. It also plays an important role in the case of the $GL(3)$-based models with  trigonometric $R$-matrix. The Izergin determinant is defined
for two sets $\bar x$ and $\bar y$ of the same cardinality $\#\bar x=\#\bar y=k$:
\begin{equation}\label{Izer}
\Izer_k(\bar x|\bar y)= \Delta'(\bar x)\Delta(\bar y)\;h(\bar x, \bar y)\;
\det_k t(x_i,y_j)\,.
\end{equation}
For further applications it is convenient to introduce two modifications of the Izergin determinant
\begin{equation}\label{Mod-Izer}
\Izerl_k(\bar x|\bar y)= \mathfrak{p}(\bar x)\cdot\Izer_k(\bar x|\bar y)\,, \qquad
\Izerr_k(\bar x|\bar y)= \mathfrak{p}(\bar y)\cdot\Izer_k(\bar x|\bar y)\,,
\end{equation}
which we call left and right Izergin determinants, respectively.

For the derivation of determinant representations for scalar products we shall use several properties of the Izergin determinant.
It is obvious that this rational function is symmetric over the set $\bar x$ and symmetric over the set $\bar y$. The Izergin determinant \eqref{Izer}
vanishes if one of its arguments goes to infinity and other arguments are fixed. Respectively, $\Izerl_k$ is bounded if $x_i\to\infty$,
while $\Izerr_k$ is bounded if $y_i\to\infty$.

Several simple properties of the Izergin determinant follow directly  from the definition \eqref{Izer}. Their detailed
proofs are given in \cite{PakRS13c}. A reduction property has the form
 \begin{equation}\label{K-red}
\Izerlr_{n+m}(\{\bar x, q^{-2}\bar z\}|\{\bar y,\bar z\}) =\Izerlr_{n+m}(\{\bar x, \bar z\}|\{\bar y,q^{2}\bar z\})=(-q)^{\mp m}\Izerlr_{n}(\bar x|\bar y),
\end{equation}
where $m=\#\bar z$ and $n=\#\bar x=\#\bar y$. Here the  superscripts $(l,r)$ on $\Izer$ mean that the equation is valid both for $\Izerl$ and for $\Izerr$ with an appropriate choice of component (first/up or second/down) throughout the equation.

A similar reduction of the Izergin determinant takes place in the poles of this function
 \begin{equation}\label{K-Res}
\Bigl.\Izerlr_{n+1}(\{\bar x,z\}|\{\bar y,z'\})\Bigr|_{z'\to z}= f(z,z')
f(z,\bar y)f(\bar x,z) \Izerlr_{n}(\bar x|\bar y)+{\rm reg},
\end{equation}
where ${\rm reg}$ means the regular part.
One more simple property of $\Izerlr_{n}$ is also useful
 \begin{equation}\label{K-invers}
  \Izerlr_{n}( q^{-2}\bar x|\bar y)=\Izerlr_{n}( \bar x|q^{2}\bar y)
= (-q)^{\mp n} f^{-1}(\bar y,\bar x) \Izerrl_{n}(\bar y|\bar x)\,.
\end{equation}

More sophisticated properties of the Izergin determinant represent  summation identities, in which the sum is taken over the partitions of one or more sets of variables.

\begin{lemma}\label{main-ident}
Let $\bar\gamma$, $\bar\alpha$ and $\bar\beta$ be three sets of complex variables with $\#\alpha=m_1$,
$\#\beta=m_2$, and $\#\gamma=m_1+m_2$. Then
\begin{equation}\label{Sym-Part-old1}
  \sum
 \Izerlr_{m_1}(\bar\gamma_{\so}|\bar \alpha)\Izerrl_{m_2}(\bar \beta|\bar\gamma_{\st})f(\bar\gamma_{\st},\bar\gamma_{\so})
 = (-q)^{\mp m_1}f(\bar\gamma,\bar \alpha) \Izerrl_{m_1+m_2}(\{\bar \alpha q^{-2},\bar \beta\}|\bar\gamma).
 \end{equation}
The sum is taken with respect to all partitions of the set $\bar\gamma\Rightarrow\{\bar\gamma_{\so},\bar\gamma_{\st}\}$ with $\#\bar\gamma_{\so}=m_1$ and $\#\bar\gamma_{\st}=m_2$.
\end{lemma}
This lemma is a complete analog of lemma~$1$ of \cite{BelPRS12b}, where the reader can find the detailed proof.

\begin{lemma}\label{Long-Det}
Let $\bar \gamma$ and $\bar\xi$ be two sets of generic complex numbers with $\#\bar\gamma=\#\bar\xi=m$. Let
also $\phi_1(\gamma)$ and $\phi_2(\gamma)$ be two arbitrary functions of a complex variable $\gamma$. Then
\begin{multline}\label{SumDet1}
\sum \Izerlr_m(\{\bar\gamma_{\so}q^{-2}, \bar\gamma_{\st}\}|\bar \xi)f(\bar \xi, \bar\gamma_{\so})f(\bar\gamma_{\st},\bar\gamma_{\so})
\phi_1(\bar\gamma_{\so})\phi_2(\bar\gamma_{\st})\num
=\mathfrak{p}^{\ell,r}\;\Delta'_m(\bar\xi)\Delta_m(\bar\gamma)
\det_m\Bigl(\phi_2(\gamma_k)t(\gamma_k,\xi_j)h(\gamma_k,\bar\xi)+q^{\mp 1}(-1)^m \phi_1(\gamma_k)t(\xi_j,\gamma_k)h(\bar\xi,\gamma_k)\Bigr).
\end{multline}
Here $\mathfrak{p}^{\ell}=\mathfrak{p}(\bar\gamma)$, $\mathfrak{p}^{r}=\mathfrak{p}(\bar\xi)$, and
we used the shorthand  notation \eqref{SH-prod} for the products of the functions $\phi_1$ and $\phi_2$. The sum is taken over all
possible partitions $\bar\gamma\Rightarrow\{\bar\gamma_{\so},\bar\gamma_{\st}\}$.
\end{lemma}
This lemma is a trigonometric analog of lemma~$2$ of \cite{BelPRS12b}. However, the proof contains small, but important differences. Therefore, we give this proof in appendix~\ref{A-Long-Det}.

\begin{lemma}\label{Wau}
Let $\bar\alpha$ and $\bar\beta$ be two sets of generic complex numbers with $\#\bar\alpha=\#\bar\beta=n$.
Let $z$ be an arbitrary complex number. Then
\begin{equation}\label{Ident-GG}
\sum  q^{n_{\so}}f(\bar\beta_{\so},z)f(\bar\beta_{\st},\bar\beta_{\so})f(\bar\alpha_{\so},\bar\alpha_{\st})   \Izerr_{n_{\so}}(\bar\beta_{\so}|\bar\alpha_{\so})\Izerl_{n_{\st}}(\bar\alpha_{\st}|\bar\beta_{\st}q^{-2})
  =q^n G_n(\bar\alpha|\bar\beta) h(\bar\alpha,z)g(\bar\beta,z),
       \end{equation}
where
\be{def G}
G_n(\bar\alpha|\bar\beta)  =(-1)^n t(\bar\alpha,\bar\beta)h(\bar\alpha,\bar\alpha)h(\bar\beta,\bar\beta),
\ee
and the sum is taken over all possible partitions $\bar\alpha\Rightarrow\{\bar\alpha_{\so},\;\bar\alpha_{\st}\}$ and $\bar\beta\Rightarrow\{\bar\beta_{\so},\;\bar\beta_{\st}\}$
with $\#\bar\alpha_{\so}=\#\bar\beta_{\so}=n_{\so}$, $n_{\so}=0,\dots,n$, and $\#\bar\alpha_{\st}=\#\bar\beta_{\st}=n_{\st}=n-n_{\so}$.
\end{lemma}
This lemma is a trigonometric analog of lemma~$A.3$ of \cite{PakRS14d}. Due to the importance of this lemma we give the proof in appendix~\ref{A-Wau}.

The identity \eqref{Ident-GG} has several particular cases.

\begin{cor}\label{Wau0}
Let $\bar\alpha$ and $\bar\beta$ be two sets of generic complex numbers with $\#\bar\alpha=\#\bar\beta=n$.
Then
\begin{equation}\label{Ident-Ginf1}
\sum f(\bar\beta_{\st},\bar\beta_{\so})f(\bar\alpha_{\so},\bar\alpha_{\st})   \Izerr_{n_{\so}}(\bar\beta_{\so}|\bar\alpha_{\so})\Izerl_{n_{\st}}(\bar\alpha_{\st}|\bar\beta_{\st}q^{-2})
=G_n(\bar\alpha|\bar\beta),
       \end{equation}
and
\begin{equation}\label{Ident-Ginf2}
\sum  q^{2n_{\st}}f(\bar\beta_{\st},\bar\beta_{\so})f(\bar\alpha_{\so},\bar\alpha_{\st})   \Izerl_{n_{\so}}(\bar\beta_{\so}|\bar\alpha_{\so})\Izerr_{n_{\st}}(\bar\alpha_{\st}|\bar\beta_{\st}q^{-2})
  =\frac{\mathfrak{p}(\bar\beta)}{\mathfrak{p}(\bar\alpha)}\;G_n(\bar\alpha|\bar\beta).
       \end{equation}
The sum is taken over all possible partitions $\bar\alpha\Rightarrow\{\bar\alpha_{\so},\;\bar\alpha_{\st}\}$ and $\bar\beta\Rightarrow\{\bar\beta_{\so},\;\bar\beta_{\st}\}$
with $\#\bar\alpha_{\so}=\#\bar\beta_{\so}=n_{\so}$, $n_{\so}=0,\dots,m$, and $\#\bar\alpha_{\st}=\#\bar\beta_{\st}=n_{\st}=n-n_{\so}$.
\end{cor}

Identity \eqref{Ident-Ginf1} follows from \eqref{Ident-GG}  in the limit $z\to\infty$.  Identity \eqref{Ident-Ginf2} follows
from \eqref{Ident-Ginf1} due to obvious relation
\be{triv-rav}
\Izerr_{n_{\so}}(\bar\beta_{\so}|\bar\alpha_{\so})\Izerl_{n_{\st}}(\bar\alpha_{\st}|\bar\beta_{\st}q^{-2})
=q^{2n_{\st}}\frac{\mathfrak{p}(\bar\alpha)}{\mathfrak{p}(\bar\beta)}\;\Izerl_{n_{\so}}(\bar\beta_{\so}|\bar\alpha_{\so})
\Izerr_{n_{\st}}(\bar\alpha_{\st}|\bar\beta_{\st}q^{-2}).
\ee

\begin{cor}\label{Wau-z0}
Under the conditions of corollary~\ref{Wau0}
\begin{equation}\label{Ident-G0-1}
\sum  q^{2n_{\so}}f(\bar\beta_{\st},\bar\beta_{\so})f(\bar\alpha_{\so},\bar\alpha_{\st})   \Izerr_{n_{\so}}(\bar\beta_{\so}|\bar\alpha_{\so})\Izerl_{n_{\st}}(\bar\alpha_{\st}|\bar\beta_{\st}q^{-2})
  = q^{2n}\;\frac{\mathfrak{p}(\bar\alpha)}{\mathfrak{p}(\bar\beta)}\;G_n(\bar\alpha|\bar\beta).
       \end{equation}
and
\begin{equation}\label{Ident-G0-2}
\sum f(\bar\beta_{\st},\bar\beta_{\so})f(\bar\alpha_{\so},\bar\alpha_{\st})   \Izerl_{n_{\so}}(\bar\beta_{\so}|\bar\alpha_{\so})\Izerr_{n_{\st}}(\bar\alpha_{\st}|\bar\beta_{\st}q^{-2})
=G_n(\bar\alpha|\bar\beta),
       \end{equation}
\end{cor}

Identity \eqref{Ident-G0-1} follows  from \eqref{Ident-GG}  in the limit $z=0$. Identity \eqref{Ident-G0-2} follows from
\eqref{Ident-G0-1} and \eqref{triv-rav}.

\begin{cor}\label{Wau0i}
Under the conditions of corollary~\ref{Wau0}
\begin{equation}\label{Ident-G0i}
\sum  q^{n_{\so}}[n_{\so}]f(\bar\beta_{\st},\bar\beta_{\so})f(\bar\alpha_{\so},\bar\alpha_{\st})   \Izerr_{n_{\so}}(\bar\beta_{\so}|\bar\alpha_{\so})\Izerl_{n_{\st}}(\bar\alpha_{\st}|\bar\beta_{\st}q^{-2})
=\frac{q^{2n}\frac{\mathfrak{p}(\bar\alpha)}{\mathfrak{p}(\bar\beta)}-1}{q-q^{-1}}
  \;G_n(\bar\alpha|\bar\beta),
       \end{equation}
where $[n_{\so}]$ means a $q$-number $[n_{\so}]=(q^{n_{\so}}-q^{-n_{\so}})/(q-q^{-1})$.
\end{cor}

This identity is a linear combination of \eqref{Ident-Ginf1} and \eqref{Ident-G0-1}.

\section{Derivation of determinant representation\label{S-DDD}}

Derivation of a determinant representation for the scalar product of twisted on-shell and usual on-shell
Bethe vectors is quite similar to the one given in \cite{BelPRS12b} for $GL(3)$-invariant models. Nevertheless, in certain
cases the $q$-deformation of the $R$-matrix leads to an essential difference in the final result. Therefore we give all the details
of the derivation.

We start with the sum formula for the scalar product in the $GL(3)$-based models with the trigonometric $R$-matrix  \cite{PakRS14b}:
\begin{multline}\label{scal}
S_{a,b}(\blac;\bmuc|\blab;\bmub)= \sum
f(\blab_{\st},\blab_{\so})f(\blac_{\so},\blac_{\st})f(\bmub_{\so},\bmub_{\st})
f(\bmuc_{\st},\bmuc_{\so})f(\bmuc_{\so},\blac_{\so}) f(\bmub_{\st},\blab_{\st})
\num
\times
\frac{r_1(\blac_\st)r_1(\blab_\so)r_3(\bmuc_\st)r_3(\bmub_\so)}{f(\bmuc,\blac)f(\bmub,\blab)}\;
\RHCl_{a-k,n}(\blac_{\st};\blab_{\st}|\bmuc_{\so};\bmub_{\so}) \;\RHCr_{k,b-n}(\blab_{\so};\blac_{\so}|\bmub_{\st};\bmuc_{\st})\;.
 \end{multline}
The sum in \eqref{scal} is taken over the partitions of the sets $\blac$, $\blab$, $\bmuc$, and $\bmub$
 \be{part-1}
 \begin{array}{ll}
 \blac\Rightarrow\{\blac_{\so},\;\blac_{\st}\}, &\qquad  \bmuc\Rightarrow\{\bmuc_{\so},\;\bmuc_{\st}\},\\
 \blab\Rightarrow\{\blab_{\so},\;\blab_{\st}\}, &\qquad  \bmub\Rightarrow\{\bmub_{\so},\;\bmub_{\st}\} .
 \end{array}
 \ee
The partitions are independent except that $\#\blab_{\so}=\#\blac_{\so}=k$ with $k=0,\dots,a$, and $\#\bmub_{\so}=\#\bmuc_{\so}=n$
with $n=0,\dots,b$.

The functions $\RHCl_{a-k,n}$ and $\RHCr_{k,b-n}$ are the left and the right highest coefficients, respectively. They can be expressed in terms
of the Izergin determinants \cite{PakRS13c}. We need two representations of these functions. The first one reads
 \begin{equation}\label{RHC-IHC-def1}
  \RHClr_{a,b}(\bar t;\bar x|\bar s;\bar y)=(-q)^{\mp b}\sum
 \Izerrl_b(\bar s|\bar w_{\so}q^2)\Izerlr_a(\bar w_{\st}|\bar t)
  \Izerlr_b(\bar y|\bar w_{\so})f(\bar w_{\so},\bar w_{\st}),
   \end{equation}
where $\bar w=\{\bar x,\bar s\}$. The sum is taken with respect to partitions of the set $\bar w\Rightarrow
\{\bar w_{\so},\bar w_{\st}\}$ with $\#\bar w_{\so}=b$ and $\#\bar w_{\st}=a$. Similarly to the formulas of section~\ref{S-SI} (see e.g.
\eqref{Sym-Part-old1}, \eqref{Long-Det})
the superscript $(l,r)$ on  ${\sf Z}_{a,b}$ means that  equation \eqref{RHC-IHC-def1} is valid for $\RHCl_{a,b}$ and for $\RHCr_{a,b}$ separately. Choosing  the first or the second component in the pair $(l,r)$ and the corresponding (up or down resp.) exponent of
$(-q)^{\mp b}$ in this equation, we obtain representations either for $\RHCl_{a,b}(\bar t;\bar x|\bar s;\bar y)$
or for $\RHCr_{a,b}(\bar t;\bar x|\bar s;\bar y)$.

The second representation for the highest coefficient has the following form:
   \begin{multline}\label{RHC-IHC-def2}
  \RHClr_{a,b}(\bar t;\bar x|\bar s;\bar y)
  =(-q)^{\mp a}f(\bar y,\bar x)f(\bar s,\bar t)\\
  \times \sum \Izerrl_a(\bar tq^{-2}|\bar\eta_{\so}q^2)\Izerlr_a(\bar xq^{-2}|\bar\eta_{\so})\Izerlr_b(\bar\eta_{\st}|\bar s)f(\bar\eta_{\so},\bar\eta_{\st}),
 \end{multline}
where $\bar\eta=\{\bar y,\bar tq^{-2}\}$. The sum is taken with respect to partitions of the set $\bar\eta\Rightarrow
\{\bar\eta_{\so},\bar\eta_{\st}\}$ with $\#\bar\eta_{\so}=a$ and $\#\bar\eta_{\st}=b$.

It is convenient to use one type of representations
for $\RHCl_{a-k,n}$ and another type for $\RHCr_{k,b-n}$. For example, using
\eqref{RHC-IHC-def1} for $\RHCl_{a-k,n}(\blac_{\st};\blab_{\st}|\bmuc_{\so};\bmub_{\so})$ we obtain
 \begin{equation}\label{RHC-IHC}
  \RHCl_{a-k,n}(\blac_{\st};\blab_{\st}|\bmuc_{\so};\bmub_{\so})=(-q)^{-n}\sum
 \Izerr_n(\bmuc_{\so}|\bar w_{\so}q^2)\Izerl_{a-k}(\bar w_{\st}|\blac_{\st})
  \Izerl_n(\bmub_{\so}|\bar w_{\so})f(\bar w_{\so},\bar w_{\st}),
 \end{equation}
with $\bar w=\{\blab_{\st},\bmuc_{\so}\}$. Thus, we obtain an additional sum over partitions of the sets $\blab_{\st}$ and $\bmuc_{\so}$
 into sub-subsets, but we do not create additional sub-subsets of $\blac$ and $\bmub$. Using now \eqref{RHC-IHC-def2} for
$\RHCr_{k,b-n}(\blab_{\so};\blac_{\so}|\bmub_{\st};\bmuc_{\st})$ we obtain
 \begin{multline}\label{Al-RHC-IHC}
 \RHCr_{k,b-n}(\blab_{\so};\blac_{\so}|\bmub_{\st};\bmuc_{\st})=(-q)^{k}f(\bmuc_{\st},\bar x)f(\bar s,\blab_{\so})\num
   \sum \Izerl_k(\blab_{\so}q^{-2}|\bar\eta_{\so}q^2)\Izerr_k(\blac_{\so}|\bar\eta_{\so}q^{2})
   \Izerr_{b-n}(\bar\eta_{\st}|\bmub_{\st})f(\bar\eta_{\so},\bar\eta_{\st}),
      \end{multline}
with $\bar\eta=\{\bmuc_{\st},\blab_{\so}q^{-2}\}$. Thus, we again create new sub-subsets of the sets $\blab$ and $\bmuc$, but we
do not touch the original partitions of the sets $\blac$ and $\bmub$. As a result we obtain a possibility to take the
sum over partitions $\blac\Rightarrow\{\blac_{\so},\;\blac_{\st}\}$ and $\bmuc\Rightarrow\{\bmuc_{\so},\;\bmuc_{\st}\}$. Indeed
substituting into \eqref{scal}: (1) representations \eqref{RHC-IHC} and \eqref{Al-RHC-IHC}; (2) twisted Bethe equations \eqref{ATEigenS-1} for the product of functions $r_1(\blac_{\st})$;  (3) standard Bethe equations \eqref{AEigenS-1} for the product of functions $r_3(\bmub_{\so})$; we obtain
\begin{multline}\label{scal-sub}
S_{a,b}(\blac;\bmuc|\blab;\bmub)= \sum (-q)^{k-n}\left(\frac{\kappa_2}{\kappa_1}\right)^{a-k}
 r_1(\blab_\so)r_3(\bmuc_\st)\,
f(\blab_{\st},\blab_{\so})f(\bmuc_{\st},\bmuc_{\so})
\num
\times  f(\blac_{\st},\blac_{\so}) f(\bmub_{\st},\bmub_{\so})
\Izerr_n(\bmuc_{\so}|\bar w_{\so}q^2)\Izerl_{a-k}(\bar w_{\st}|\blac_{\st})
  \Izerl_n(\bmub_{\so}|\bar w_{\so})f(\bar w_{\so},\bar w_{\st})\num
  \times \Izerl_k(\blab_{\so}q^{-2}|\bar\eta_{\so}q^2)\Izerr_k(\blac_{\so}|\bar\eta_{\so}q^{2})
   \Izerr_{b-n}(\bar\eta_{\st}|\bmub_{\st})f(\bar\eta_{\so},\bar\eta_{\st})\;.
 \end{multline}
Here the sum is taken over the partitions \eqref{part-1} and additional partitions
\be{part-1a}
\begin{aligned}
\{\blab_{\st},\bmuc_{\so}\}&=\bar w\Rightarrow \{\bar w_{\so},\bar w_{\st}\},\\
\{\bmuc_{\st},\blab_{\so}q^{-2}\}&=\bar\eta\Rightarrow \{\bar\eta_{\so},\bar\eta_{\st}\}.
\end{aligned}
\ee

Now we can apply  lemma~\ref{main-ident} for the summation over the partitions  $\blac\Rightarrow\{\blac_{\so},\;\blac_{\st}\}$ and $\bmuc\Rightarrow\{\bmuc_{\so},\;\bmuc_{\st}\}$:
\be{ML-1}
\sum\Izerr_k(\blac_{\so}|\bar\eta_{\so}q^{2})\Izerl_{a-k}(\bar w_{\st}|\blac_{\st})f(\blac_{\st},\blac_{\so})
=(-q)^k   f^{-1}(\bar\eta_{\so},\blac)\,\Izerl_{a}(\{\bar\eta_{\so},\bar w_{\st}\}|\blac),
\ee
%
\be{ML-2}
\sum\Izerl_n(\bmub_{\so}|\bar w_{\so})\Izerr_{b-n}(\bar\eta_{\st}|\bmub_{\st})f(\bmub_{\st},\bmub_{\so})
=(-q)^{-n}\,f(\bmub,\bar w_{\so})\Izerr_{b}(\{\bar w_{\so}q^{-2},\bar\eta_{\st}\}|\bmub),
\ee
where we used \eqref{formulas}. Then we arrive at
\begin{multline}\label{scal-step1}
S_{a,b}(\blac;\bmuc|\blab;\bmub)= \sum (-q)^{2(k-n)}\left(\frac{\kappa_2}{\kappa_1}\right)^{a-k}
r_1(\blab_\so)r_3(\bmuc_\st)\num
\times
f(\blab_{\st},\blab_{\so})f(\bmuc_{\st},\bmuc_{\so})\frac{f(\bmub,\bar w_{\so})}{f(\bar\eta_{\so},\blac)}
f(\bar w_{\so},\bar w_{\st})f(\bar\eta_{\so},\bar\eta_{\st})\num
\times \Izerr_n(\bmuc_{\so}|\bar w_{\so}q^2)\Izerl_k(\blab_{\so}q^{-2}|\bar\eta_{\so}q^2)
  %
  \Izerl_{a}(\{\bar\eta_{\so},\bar w_{\st}\}|\blac) \Izerr_{b}(\{\bar w_{\so}q^{-2},\bar\eta_{\st}\}|\bmub)\;.
 \end{multline}

\subsection{Summation of sub-partitions}

In order to make the next step in the summation over the partitions we should specify the subsets of $\bar w$ and $\bar\eta$.
For this we introduce new sub-subsets as follows:
\be{New-part2}
\begin{aligned}
&\bar\eta_{\so}\Rightarrow\{\blab_{\rm ii}q^{-2},\bmuc_{\rm i}\},\qquad &\blab_{\so}\Rightarrow\{\blab_{\rm i},\blab_{\rm ii}\},\\
&\bar\eta_{\st}\Rightarrow\{\blab_{\rm i}q^{-2},\bmuc_{\rm ii}\},\qquad &\blab_{\st}\Rightarrow\{\blab_{\rm iii},\blab_{\rm iv}\},\\
&{}&{}\\
&\bar w_{\so}\Rightarrow\{\blab_{\rm iv},\bmuc_{\rm iii}\},\qquad &\bmuc_{\so}\Rightarrow\{\bmuc_{\rm iii},\bmuc_{\rm iv}\},\\
&\bar w_{\st}\Rightarrow\{\blab_{\rm iii},\bmuc_{\rm iv}\},\qquad &\bmuc_{\st}\Rightarrow\{\bmuc_{\rm i},\bmuc_{\rm ii}\}.
\end{aligned}
\ee
The cardinalities of the sub-subsets above are $\#\blab_j=k_j$ and $\#\bmuc_j=n_j$. Evidently
$\sum_{j={\rm i}}^{\rm iv}k_j=a$, $\sum_{j={\rm i}}^{\rm iv}n_j=b$ and one can  easily  see that
\be{card-new-part2}
\begin{aligned}
&n_{\rm i}+n_{\rm ii}=b-n,\qquad &k_{\rm i}+k_{\rm ii}=k,\\
&n_{\rm iii}+n_{\rm iv}=n,\qquad &k_{\rm iii}+k_{\rm iv}=a-k,\\
&n_{\rm i}=k_{\rm i},\qquad &n_{\rm iv}=k_{\rm iv}.
\end{aligned}
\ee
{\sl Remark.} Note that some relations between the cardinalities of the sub-subsets are implicitly shown by the subscripts of the Izergin determinants.
For instance, equation \eqref{scal-step1} contains the Izergin determinant $\Izerr_n(\bmuc_{\so}|\bar w_{\so}q^2)$. This means that
$\#\bmuc_{\so}=\#\bar w_{\so}=n$.  Using the fact that $\bmuc_{\so}=\{\bmuc_{\rm iii},\bmuc_{\rm iv}\}$ and $\bar w_{\so}=\{\blab_{\rm iv},\bmuc_{\rm iii}\}$
we conclude that $n_{\rm iii}+n_{\rm iv}=n_{\rm iii}+k_{\rm iv}=n$. From this we find $n_{\rm iv}=k_{\rm iv}$.

Using the new sub-subsets we recast \eqref{scal-step1} as follows:
 \begin{multline}\label{sum-New-ss2}
S_{a,b}=\sum \left(\frac{\kappa_2}{\kappa_1}\right)^{a-k}
(-q)^{2(k_{\rm i}+k_{\rm ii}-n_{\rm iii}-n_{\rm iv})}
  r_1(\blab_{\rm i})r_1(\blab_{\rm ii})r_3(\bmuc_{\rm i})r_3(\bmuc_{\rm ii})
\num
 \times f(\blab_{\rm iii},\blab_{\rm i})f(\blab_{\rm iii},\blab_{\rm ii})f(\blab_{\rm iv},\blab_{\rm i})
  f(\blab_{\rm iv},\blab_{\rm ii})
 f(\blab_{\rm iv},\blab_{\rm iii})f(\blab_{\rm ii},\blab_{\rm i})\;
 f(\bmuc_{\rm i},\bmuc_{\rm iii})f(\bmuc_{\rm i},\bmuc_{\rm iv})\num
\times f(\bmuc_{\rm ii},\bmuc_{\rm iii}) f(\bmuc_{\rm ii},\bmuc_{\rm iv})
 f(\bmuc_{\rm i},\bmuc_{\rm ii}) f(\bmuc_{\rm iii},\bmuc_{\rm iv})\;
 f(\blab_{\rm iv},\bmuc_{\rm iv})f(\bmuc_{\rm iii},\blab_{\rm iii})
 \frac{f(\bmuc_{\rm i},\blab_{\rm i}q^{-2})}{f(\bmuc_{\rm ii},\blab_{\rm ii})}\num
  \times  \frac{f(\blac,\blab_{\rm ii})f(\bmub,\blab_{\rm iv}) f(\bmub,\bmuc_{\rm iii})}{f(\bmuc_{\rm i},\blac)}\;
   \Izerr_{n}(\{\bmuc_{\rm iii},\bmuc_{\rm iv}\}|\{\blab_{\rm iv}q^2,\bmuc_{\rm iii}q^2\})\;
   \Izerl_{k}(\{\blab_{\rm i}q^{-2},\blab_{\rm ii}q^{-2}\}|\{\blab_{\rm ii},\bmuc_{\rm i}q^2\})\num
  \times
  \Izerl_{a}(\{\blab_{\rm ii}q^{-2},\blab_{\rm iii}, \bmuc_{\rm i},\bmuc_{\rm iv}\}|\blac)\;
  \Izerr_{b}(\{ \blab_{\rm i}q^{-2},\blab_{\rm iv}q^{-2},\bmuc_{\rm iii}q^{-2},\bmuc_{\rm ii}\}|\bmub)\,.
 \end{multline}
Here we used $f(xq^{-2},y)=f(x,yq^{2})=f^{-1}(y,x)$.

The next several steps are relatively simple transformations of \eqref{sum-New-ss2}. First of all
we combine  sub-subsets \eqref{New-part2} into new groups
\be{NNew-part}
\begin{aligned}
&\{\blab_{\rm i},\blab_{\rm iv}\}=\blab_{\so},\qquad &\{\bmuc_{\rm i},\bmuc_{\rm iv}\}=\bmuc_{\so},\\
&\{\blab_{\rm ii},\blab_{\rm iii}\}=\blab_{\st},\qquad &\{\bmuc_{\rm ii},\bmuc_{\rm iii}\}=\bmuc_{\st},
\end{aligned}
\ee
and  denote $n_{\so}=\#\blab_{\so}=\#\bmuc_{\so}$.  Pay attention that the subsets $\blab_{\so}$, $\blab_{\st}$,
$\bmuc_{\so}$, and $\bmuc_{\st}$ are not the same as in \eqref{scal-step1}. We use, however, the
same notation, as we are dealing with the sum over partitions, and therefore it does not matter how we denote separate terms
of this sum.

The sum in \eqref{sum-New-ss2}  takes the form
 \begin{multline}\label{sum-New-ss3}
S_{a,b}=\sum \left(\frac{\kappa_2}{\kappa_1}\right)^{a-k}
(-q)^{2(k_{\rm i}+k_{\rm ii}-n_{\rm iii}-n_{\rm iv})}
  r_1(\blab_{\rm i})r_1(\blab_{\rm ii})r_3(\bmuc_{\rm i})r_3(\bmuc_{\rm ii})
\num
 \times f(\blab_{\rm iii},\blab_{\rm ii})f(\blab_{\rm iv},\blab_{\rm i})
f(\blab_{\rm iv},\blab_{\st}) f(\blab_{\st},\blab_{\rm i})\;
f(\bmuc_{\rm i},\bmuc_{\st})f(\bmuc_{\st},\bmuc_{\rm iv})f(\bmuc_{\rm i},\bmuc_{\rm iv})f(\bmuc_{\rm ii},\bmuc_{\rm iii})
\num
\times
 f(\blab_{\rm iv},\bmuc_{\rm iv})f(\bmuc_{\rm iii},\blab_{\rm iii})
 \frac{f(\bmuc_{\rm i},\blab_{\rm i}q^{-2})f(\blac,\blab_{\rm ii})f(\bmub,\blab_{\rm iv}) f(\bmub,\bmuc_{\rm iii})}
 {f(\bmuc_{\rm ii},\blab_{\rm ii})f(\bmuc_{\rm i},\blac)}\num
  \times
   \Izerr_{n}(\{\bmuc_{\rm iii},\bmuc_{\rm iv}\}|\{\blab_{\rm iv}q^2,\bmuc_{\rm iii}q^2\})\;
   \Izerl_{k}(\{\blab_{\rm i}q^{-2},\blab_{\rm ii}q^{-2}\}|\{\blab_{\rm ii},\bmuc_{\rm i}q^2\})\num
  \times
  \Izerl_{a}(\{\blab_{\rm ii}q^{-2},\blab_{\rm iii}, \bmuc_{\so}\}|\blac)\;
  \Izerr_{b}(\{  \blab_{\so}q^{-2},\bmuc_{\rm iii}q^{-2},\bmuc_{\rm ii}\}|\bmub)\,.
 \end{multline}
Simplifying the Izergin determinants $\Izerr_{n}$ and $\Izerl_{k}$ via \eqref{K-red} and \eqref{K-invers} we have
\be{simpl-K1}
\Izerr_{n}(\{\bmuc_{\rm iii},\bmuc_{\rm iv}\}|\{\blab_{\rm iv}q^2,\bmuc_{\rm iii}q^2\})=
(-q)^{n_{\rm iii}+n_{\rm iv}} f^{-1}(\blab_{\rm iv},\bmuc_{\rm iv}) \Izerl_{n_{\rm iv}}(\blab_{\rm iv}|\bmuc_{\rm iv})\,,
\ee
and
\be{simpl-K2}
\Izerl_{k}(\{\blab_{\rm i}q^{-2},\blab_{\rm ii}q^{-2}\}|\{\blab_{\rm ii},\bmuc_{\rm i}q^2\})=
(-q)^{-(k_{\rm i}+k_{\rm ii})} f^{-1}(\bmuc_{\rm i},\blab_{\rm i}q^{-2})
\Izerr_{n_{\rm i}}(\bmuc_{\rm i}|\blab_{\rm i}q^{-2})\,.
\ee
After substitution these formulas into \eqref{sum-New-ss3} we arrive at
 \begin{multline}\label{sum-New-ss4}
S_{a,b}=\sum \left(\frac{\kappa_2}{\kappa_1}\right)^{a-k_{\rm i}-k_{\rm ii}}
(-q)^{k_{\rm i}+k_{\rm ii}-n_{\rm iii}-n_{\rm iv}}
  r_1(\blab_{\rm i})r_1(\blab_{\rm ii})r_3(\bmuc_{\rm i})r_3(\bmuc_{\rm ii})
\num
 \times f(\blab_{\rm iii},\blab_{\rm ii})f(\blab_{\rm iv},\blab_{\rm i})
f(\blab_{\rm iv},\blab_{\st}) f(\blab_{\st},\blab_{\rm i})\;
f(\bmuc_{\rm i},\bmuc_{\st})f(\bmuc_{\st},\bmuc_{\rm iv})f(\bmuc_{\rm i},\bmuc_{\rm iv})f(\bmuc_{\rm ii},\bmuc_{\rm iii})
\num
\times
  \frac{f(\bmuc_{\rm iii},\blab_{\rm iii})f(\blac,\blab_{\rm ii})f(\bmub,\blab_{\rm iv}) f(\bmub,\bmuc_{\rm iii})}
 {f(\bmuc_{\rm ii},\blab_{\rm ii})f(\bmuc_{\rm i},\blac)}\;
\Izerl_{n_{\rm iv}}(\blab_{\rm iv}|\bmuc_{\rm iv})\;
\Izerr_{n_{\rm i}}(\bmuc_{\rm i}|\blab_{\rm i}q^{-2})\num
  \times
  \Izerl_{a}(\{\blab_{\rm ii}q^{-2},\blab_{\rm iii},\bmuc_{\so}|\blac)\;
  \Izerr_{b}(\{ \blab_{\so}q^{-2},\bmuc_{\rm iii}q^{-2},\bmuc_{\rm ii}\}|\bmub)\,.
 \end{multline}

The last preliminary step is to introduce two auxiliary functions
\be{aux-f}
\hat r_1(\lab_k)= \frac{r_1(\lab_k)f(\blab_k,\lab_k)}{f(\lab_k,\blab_k)f(\bmub,\lab_k)}, \qquad
\hat r_3(\muc_k)= \frac{\kappa_2r_3(\muc_k)f(\muc_k,\bmuc_k)}{\kappa_3 f(\bmuc_k,\muc_k) f(\muc_k,\blac)}.
\ee
Observe that if $\lab_k$ satisfies Bethe equations \eqref{AEigenS-1} and $\muc_k$ satisfies twisted Bethe equations \eqref{ATEigenS-1},
then $\hat r_1(\lab_k)=\hat r_3(\muc_k)=1$. However, we do not use this fact, treating up to some stage the sets $\blab$ and $\bmuc$
as generic complex numbers. The reason of this will be clear very soon.

Then one can write the products $r_1(\blab_{\rm i})$ and $r_3(\bmuc_{\rm i})$ as follows:
\be{r1-BE}
r_1(\blab_{\rm i})= \hat r_1(\blab_{\rm i})
\frac{f(\blab_{\rm i},\blab_{\st})f(\blab_{\rm i},\blab_{\rm iv})}
{f(\blab_{\st},\blab_{\rm i})f(\blab_{\rm iv},\blab_{\rm i})}f(\bmub,\blab_{\rm i}),
\ee
\be{r3-BE}
r_3(\bmuc_{\rm i})=\hat r_3(\bmuc_{\rm i})\left(\frac{\kappa_2}{\kappa_3}\right)^{n_{\rm i}}
\frac{f(\bmuc_{\st},\bmuc_{\rm i})f(\bmuc_{\rm iv},\bmuc_{\rm i})}
{f(\bmuc_{\rm i},\bmuc_{\st})f(\bmuc_{\rm i},\bmuc_{\rm iv})}
f(\bmuc_{\rm i},\blac).
\ee
Substituting \eqref{r1-BE} and \eqref{r3-BE} into \eqref{sum-New-ss4} we obtain
 \begin{multline}\label{sum-New-ss5}
S_{a,b}=\sum \left(\frac{\kappa_2}{\kappa_1}\right)^{a-k_{\rm ii}}
\left(\frac{\kappa_1}{\kappa_3}\right)^{n_{\rm i}}
(-q)^{k_{\rm i}+k_{\rm ii}-n_{\rm iii}-n_{\rm iv}}
f(\blab_{\so},\blab_{\st}) f(\bmuc_{\st},\bmuc_{\so})\num
 \times \hat r_1(\blab_{\rm i})\hat r_3(\bmuc_{\rm i})
f(\blab_{\rm i},\blab_{\rm iv})f(\bmuc_{\rm iv},\bmuc_{\rm i})\Izerl_{n_{\rm iv}}(\blab_{\rm iv}|\bmuc_{\rm iv})\;
\Izerr_{n_{\rm i}}(\bmuc_{\rm i}|\blab_{\rm i}q^{-2})\;\frac{f(\bmuc_{\rm iii},\blab_{\rm iii})}
 {f(\bmuc_{\rm ii},\blab_{\rm ii})}
\num
  \times\Izerr_{b}(\{ \blab_{\so}q^{-2},\bmuc_{\rm iii}q^{-2},\bmuc_{\rm ii}\}|\bmub)r_3(\bmuc_{\rm ii})
  f(\bmuc_{\rm ii},\bmuc_{\rm iii})f(\bmub,\blab_{\rm iv}) f(\bmub,\bmuc_{\rm iii})  \num
  \times \Izerl_{a}(\{\blab_{\rm ii}q^{-2},\blab_{\rm iii},\bmuc_{\so}|\blac)\;r_1(\blab_{\rm ii})
  f(\blab_{\rm iii},\blab_{\rm ii})f(\blac,\blab_{\rm ii})\,.
 \end{multline}

Our goal is to take the sum over partitions $\blab_{\st}\Rightarrow\{\blab_{\rm ii},\blab_{\rm iii}\}$ and
$\bmuc_{\st}\Rightarrow\{\bmuc_{\rm ii},\bmuc_{\rm iii}\}$ via lemma~\ref{Long-Det}. For this purpose we present the
ratio $f(\bmuc_{\rm iii},\blab_{\rm iii})/f(\bmuc_{\rm ii},\blab_{\rm ii})$ in the following form:
\be{Prosto}
\frac{f(\bmuc_{\rm iii},\blab_{\rm iii})}{f(\bmuc_{\rm ii},\blab_{\rm ii})}=
\frac{f(\bmuc_{\st},\blab_{\st})}{f(\bmuc_{\st},\blab_{\rm ii})f(\bmuc_{\rm ii},\blab_{\st})}
=f(\bmuc_{\st},\blab_{\st})\,\frac{f(\bmuc_{\so},\blab_{\rm ii})f(\bmuc_{\rm ii},\blab_{\so})}{f(\bmuc,\blab_{\rm ii})f(\bmuc_{\rm ii},\blab)}\;.
\ee
Then \eqref{sum-New-ss5} turns into
 \begin{multline}\label{sum-New-ss6}
S_{a,b}=\sum
\left(\frac{\kappa_1}{\kappa_3}\right)^{n_{\rm i}}(-q)^{2n_{\rm i}}
  f(\bmuc_{\st},\blab_{\st})f(\blab_{\so},\blab_{\st}) f(\bmuc_{\st},\bmuc_{\so})\num
 \times
\hat r_1(\blab_{\rm i})\hat r_3(\bmuc_{\rm i})f(\blab_{\rm i},\blab_{\rm iv})f(\bmuc_{\rm iv},\bmuc_{\rm i})\Izerl_{n_{\rm iv}}(\blab_{\rm iv}|\bmuc_{\rm iv})\;
\Izerr_{n_{\rm i}}(\bmuc_{\rm i}|\blab_{\rm i}q^{-2})
\num
  \times\Izerr_{b}(\{ \blab_{\so}q^{-2},\bmuc_{\rm iii}q^{-2},\bmuc_{\rm ii}\}|\bmub)
  \frac{(-q)^{n_{\rm ii}-b}r_3(\bmuc_{\rm ii})}  {f(\bmuc_{\rm ii},\blab)}
    f(\bmuc_{\rm ii},\bmuc_{\rm iii})f(\bmub,\blab_{\rm iv}) f(\bmub,\bmuc_{\rm iii})
  f(\bmuc_{\rm ii},\blab_{\so})  \num
  \times \Izerl_{a}(\{\blab_{\rm ii}q^{-2},\blab_{\rm iii},\bmuc_{\so}|\blac)\;
  \frac{(-q)^{k_{\rm ii}}r_1(\blab_{\rm ii})}  {f(\bmuc,\blab_{\rm ii})}
  \left(\frac{\kappa_2}{\kappa_1}\right)^{a-k_{\rm ii}}
  f(\blab_{\rm iii},\blab_{\rm ii})f(\blac,\blab_{\rm ii})f(\bmuc_{\so},\blab_{\rm ii})\,,
 \end{multline}
where we have used $k_{\rm i}=n_{\rm i}$.

The sum over partitions $\blab_{\st}\Rightarrow\{\blab_{\rm ii},\blab_{\rm iii}\}$ is only in the
last line of \eqref{sum-New-ss6}. We can apply  lemma~\ref{Long-Det} to this sum if we set
\be{c1c22}
\phi_1(\gamma)= -q\frac{r_1(\gamma)}{f(\bmuc,\gamma)}, \qquad \phi_2(\gamma)=\frac{\kappa_2}{\kappa_1}\;.
\ee
Consider the application of lemma~\ref{Long-Det} in more detail. Denote the sum over the partitions
$\blab_{\st}\Rightarrow\{\blab_{\rm ii},\blab_{\rm iii}\}$ in the last line of \eqref{sum-New-ss6} by $U$:
\be{U1}
U=\sum_{\blab_{\st}\Rightarrow\{\blab_{\rm ii},\blab_{\rm iii}\}}
\Izerl_{a}(\{\blab_{\rm ii}q^{-2},\blab_{\rm iii},\bmuc_{\so}|\blac)\;
  \frac{(-q)^{k_{\rm ii}}r_1(\blab_{\rm ii})}  {f(\bmuc,\blab_{\rm ii})}
  \left(\frac{\kappa_2}{\kappa_1}\right)^{a-k_{\rm ii}}
  f(\blab_{\rm iii},\blab_{\rm ii})f(\blac,\blab_{\rm ii})f(\bmuc_{\so},\blab_{\rm ii}).
\ee
Let us also write \eqref{SumDet1} with the functions $\phi_1$ and $\phi_2$ as in \eqref{c1c22} and $m=a$
\begin{equation}\label{SumDet2}
\mathfrak{p}(\bar\gamma){\sf L}_{a}(\bar\gamma|\bar\xi)=\sum_{\bar\gamma\Rightarrow\{\bar\gamma_{\so},\bar\gamma_{\st}\} }
\Izerl_a(\{\bar\gamma_{\so}q^{-2}, \bar\gamma_{\st}\}|\bar \xi)
(-q)^{m_{\so}}\left(\frac{\kappa_2}{\kappa_1}\right)^{m_{\st}} \frac{r_1(\bar\gamma_{\so})}{f(\bmuc,\bar\gamma_{\so})}
f(\bar \xi, \bar\gamma_{\so})f(\bar\gamma_{\st},\bar\gamma_{\so}),
\end{equation}
where $m_{\so}=\#\bar\gamma_{\so}$,  $m_{\st}=\#\bar\gamma_{\st}$, and
\be{L-def}
{\sf L}_{a}(\bar\gamma|\bar\xi)=\Delta'_a(\bar\xi)\Delta_a(\bar\gamma)\;
\det_a\bigl[{\sf N}^{(u)}(\xi_j,\gamma_k)h(\gamma_k,\blab)\bigr],
\ee
with ${\sf N}^{(u)}(\xi_j,\gamma_k)$ defined in \eqref{Nu-def}.

Let now $\bar\gamma=\{\blab_{\st},\bmuc_{\so}\}$.  Observe that $\phi_1(\muc_k)=0$ due to \eqref{c1c22}. Therefore
if $\bmuc_{\so}\cap\gamma_{\so}\ne\emptyset$, then the corresponding contribution to \eqref{SumDet2} vanishes. Hence, all the
non-vanishing contributions to the sum  \eqref{SumDet2} correspond only to such partitions for which $\bmuc_{\so}\subset\gamma_{\st}$.
Therefore, we can set $\gamma_{\so}=\blac_{\rm ii}$ and $\gamma_{\st}=\{\blac_{\rm iii},\bmuc_{\so}\}$. Then $m_{\so}=k_{\rm ii}$,
$m_{\st}=a-k_{\rm ii}$, and
\eqref{SumDet2} turns into
\begin{multline}\label{U2}
\mathfrak{p}(\blab_{\st})\mathfrak{p}(\bmuc_{\so})
{\sf L}_{a}(\{\blab_{\st},\bmuc_{\so}\}|\bar\xi)=\sum_{\blab_{\st}\Rightarrow\{\blab_{\rm ii},\blab_{\rm iii}\}}
\Izerl_{a}(\{\blab_{\rm ii}q^{-2},\blab_{\rm iii},\bmuc_{\so}|\bar\xi)\\
 \times  \frac{(-q)^{k_{\rm ii}}r_1(\blab_{\rm ii})}  {f(\bmuc,\blab_{\rm ii})}
  \left(\frac{\kappa_2}{\kappa_1}\right)^{a-k_{\rm ii}}
  f(\blab_{\rm iii},\blab_{\rm ii})f(\bar\xi,\blab_{\rm ii})f(\bmuc_{\so},\blab_{\rm ii}).
\end{multline}
It remains to set $\bar\xi=\blac$, and we reproduce \eqref{U1}. Thus, $U=
\mathfrak{p}(\blab_{\st})\mathfrak{p}(\bmuc_{\so}){\sf L}_{a}(\{\blab_{\st},\bmuc_{\so}\}|\blac)$.

Similarly, the sum over partitions $\bmuc_{\st}\Rightarrow\{\bmuc_{\rm ii},\bmuc_{\rm iii}\}$ is only in the
third line of \eqref{sum-New-ss6}. We can apply again  lemma~\ref{Long-Det} to this sum with the
following identifications:
\be{c1c21}
\phi_1(\gamma)=-q^{-1}, \qquad \phi_2(\gamma)= \frac{r_3(\gamma)}{f(\gamma,\blab)}\;.
\ee

Thus, we obtain
 \begin{multline}\label{Part-sum-1}
S_{a,b}=\sum
  f(\bmuc_{\st},\blab_{\st})f(\blab_{\so},\blab_{\st}) f(\bmuc_{\st},\bmuc_{\so})\mathfrak{p}(\bmub)
  \mathfrak{p}(\blab_{\st})\mathfrak{p}(\bmuc_{\so})\\
  \times {\sf G}^{(\kappa)}_{n_{\so}}(\blab_{\so},\bmuc_{\so})
  {\sf L}_{a}(\{\blab_{\st},\bmuc_{\so}\}|\blac) {\sf M}_{b}(\{\blab_{\so},\bmuc_{\st}\}|\bmub),
  \end{multline}
where
\be{M-def}
{\sf M}_{b}(\bar\gamma|\bmub)=\Delta'_b(\bmub)\Delta_b(\bar\gamma)\;
\det_b\bigl[{\sf N}^{(v)}(\mub_j,\gamma_k)h(\bmuc,\gamma_k)\bigr],
\ee
with ${\sf N}^{(v)}(\mub_j,\gamma_k)$ defined in \eqref{Nv-def}.

The function ${\sf G}^{(\kappa)}_{n_{\so}}$ is still given as the sum over the partitions
$\blab_{\so}\Rightarrow\{\blab_{\rm i},\blab_{\rm iv}\}$ and $\bmuc_{\so}\Rightarrow\{\bmuc_{\rm i},\bmuc_{\rm iv}\}$:
\be{G-kappa}
   {\sf G}^{(\kappa)}_{n_{\so}}(\blab_{\so},\bmuc_{\so})=\sum
 \left(\frac{\kappa_1}{\kappa_3}\right)^{n_{\rm i}}q^{2n_{\rm i}}\hat r_1(\blab_{\rm i})\hat r_3(\bmuc_{\rm i})
f(\blab_{\rm i},\blab_{\rm iv})f(\bmuc_{\rm iv},\bmuc_{\rm i})\Izerl_{n_{\rm iv}}(\blab_{\rm iv}|\bmuc_{\rm iv})\;
\Izerr_{n_{\rm i}}(\bmuc_{\rm i}|\blab_{\rm i}q^{-2}).
\ee

It is important that up to now we did not use (twisted) Bethe equations for the parameters $\bmuc$ and $\blab$. In all the above calculations  we considered these parameters as generic complex numbers. Therefore, the matrix elements
${\sf N}^{(u)}(\lac_j,\lab_k)$ \eqref{Nu-def} and ${\sf N}^{(v)}(\mub_j,\muc_k)$ \eqref{Nv-def} are well defined even in the case when
$\blac\cap\blab\ne\emptyset$ or $\bmuc\cap\bmub\ne\emptyset$. Indeed, since we have no restriction for $\bmuc$ and $\blab$, it is no
problem to take the limits $\lab_k\to\lac_j$ in \eqref{Nu-def} or $\muc_k\to\mub_j$ in \eqref{Nv-def}. Therefore, if some of the Bethe parameters
from different Bethe vectors coincide, then the corresponding matrix elements ${\sf N}^{(u)}(\lac_j,\lab_k)$ and ${\sf N}^{(v)}(\mub_j,\muc_k)$
should be understood as in \eqref{diag-el1} and \eqref{diag-el2}.

\section{Final summations\label{S-FS}}

After we have defined the matrices ${\sf N}^{(u)}$ and ${\sf N}^{(v)}$ for all possible values of the Bethe parameters, we can impose
(twisted) Bethe equations for the sets $\bmuc$ and $\blab$. Then $\hat r_1(\lab_k)\hat r_3(\muc_k)=1$ and the sum in \eqref{G-kappa}  turns into
\be{G-kappa1}
   {\sf G}^{(\kappa)}_{n_{\so}}(\blab_{\so},\bmuc_{\so})=\sum
 \left(\frac{\kappa_1}{\kappa_3}\right)^{n_{\rm i}}q^{2n_{\rm i}}f(\blab_{\rm i},\blab_{\rm iv})
 f(\bmuc_{\rm iv},\bmuc_{\rm i})\Izerl_{n_{\rm iv}}(\blab_{\rm iv}|\bmuc_{\rm iv})\;
\Izerr_{n_{\rm i}}(\bmuc_{\rm i}|\blab_{\rm i}q^{-2}).
\ee
We cannot calculate this sum at generic $\kappa_1/\kappa_3$. However,
we know the result for ${\sf G}^{(\kappa)}_{n_{\so}}$ in two particular cases, namely $\kappa_3/\kappa_1=1$ and $\kappa_3/\kappa_1=q^2$. They are
described by corollaries~\ref{Wau0} and~\ref{Wau-z0}. Let us denote the corresponding functions ${\sf G}^{(\kappa)}_{n_{\so}}$
by ${\sf G}^{(1)}_{n_{\so}}$ and ${\sf G}^{(q^2)}_{n_{\so}}$, respectively. Then we have
\be{G1}
{\sf G}^{(1)}_{n_{\so}}=(-1)^{n_{\so}} \frac{\mathfrak{p}(\blab_{\so})}{\mathfrak{p}(\bmuc_{\so})}\;
t(\bmuc_{\so},\blab_{\so})h(\bmuc_{\so},\bmuc_{\so})h(\blab_{\so},\blab_{\so}),
\ee
and
\be{Gq2}
{\sf G}^{(q^2)}_{n_{\so}}=(-1)^{n_{\so}}\;
t(\bmuc_{\so},\blab_{\so})h(\bmuc_{\so},\bmuc_{\so})h(\blab_{\so},\blab_{\so}).
\ee
Consider these two cases separately.

\subsection{The case $\kappa_3/\kappa_1=1$}

Without loss of generality we can set $\kappa_1=\kappa_3=1$. Let us denote the corresponding
scalar product by $S^{(1)}_{a,b}$. Substituting \eqref{G1} into \eqref{Part-sum-1} we obtain
 \begin{multline}\label{Part-sum-2}
S^{(1)}_{a,b}=\mathfrak{p}(\bmub)\mathfrak{p}(\blab)\sum (-1)^{n_{\so}}
  f(\bmuc_{\st},\blab_{\st})f(\blab_{\so},\blab_{\st}) f(\bmuc_{\st},\bmuc_{\so})
  t(\bmuc_{\so},\blab_{\so})h(\blab_{\so},\blab_{\so})h(\bmuc_{\so},\bmuc_{\so})
\num
 \times  {\sf L}_{a}(\{\blab_{\st},\bmuc_{\so}\}|\blac) {\sf M}_{b}(\{\blab_{\so},\bmuc_{\st}\}|\bmub).
\end{multline}

Now we introduce
\be{hLhM}
\hat{\sf L}_{a}(\bar\gamma|\blac)=\frac{{\sf L}_{a}(\bar\gamma|\blac)}{h(\bar\gamma,\blab)},\qquad
\hat{\sf M}_{b}(\bar\gamma|\bmub)=\frac{{\sf M}_{b}(\bar\gamma|\bmub)}{h(\bmuc,\bar\gamma)},
\ee
and then the equation \eqref{Part-sum-2} takes the form
 \begin{multline}\label{Part-sum-3}
S^{(1)}_{a,b}= \mathfrak{p}(\bmub)\mathfrak{p}(\blab)\sum
(-1)^{n_{\so}}f(\bmuc_{\st},\blab_{\st})f(\blab_{\so},\blab_{\st}) f(\bmuc_{\st},\bmuc_{\so})
  t(\bmuc_{\so},\blab_{\so})h(\blab_{\so},\blab_{\so})h(\bmuc_{\so},\bmuc_{\so})
\num
 \times  h(\blab_{\st},\blab)h(\bmuc_{\so},\blab)   h(\bmuc,\blab_{\so})h(\bmuc,\bmuc_{\st})
 \hat{\sf L}_{a}(\{\blab_{\st},\bmuc_{\so}\}|\blac) \hat{\sf M}_{b}(\{\blab_{\so},\bmuc_{\st}\}|\bmub).
\end{multline}
It convenient to present all the functions $f(x,y)$ as $f(x,y)=g(x,y)h(x,y)$, all the functions $t(x,y)$ as
$t(x,y)=g(x,y)/h(x,y)$, and then to combine separately all the functions $h(x,y)$  and all the functions $g(x,y)$.
We obtain
 \begin{multline}\label{Part-sum-4}
S^{(1)}_{a,b}= \mathfrak{p}(\bmub)\mathfrak{p}(\blab)C_h\sum
g(\bmuc_{\st},\blab_{\st})g(\blab_{\so},\blab_{\st}) g(\bmuc_{\st},\bmuc_{\so})
  g(\blab_{\so},\bmuc_{\so})
\num
 \times
 \hat{\sf L}_{a}(\{\blab_{\st},\bmuc_{\so}\}|\blac) \hat{\sf M}_{b}(\{\blab_{\so},\bmuc_{\st}\}|\bmub),
\end{multline}
where $C_h$ is given by \eqref{Ch}.
Recall also that
\be{hLM-def}
\hat{\sf L}_{a}(\bar\gamma|\blac)=\Delta'_a(\blac)\Delta_a(\bar\gamma)\;
\det_a{\sf N}^{(u)}(\lac_j,\gamma_k),\qquad
\hat{\sf M}_{b}(\bar\gamma|\bmub)=\Delta'_b(\bmub)\Delta_b(\bar\gamma)\;
\det_b{\sf N}^{(v)}(\mub_j,\gamma_k),
\ee
and the matrices ${\sf N}^{(u)}(\lac_j,\gamma_k)$, ${\sf N}^{(v)}(\mub_j,\gamma_k)$ are given by
\eqref{Nu-def}, \eqref{Nv-def}.
It remains to apply the following

\begin{prop}\label{GenMat}
Let ${\sf N}$ be an $(a+b)\times (a+b)$ matrix and $\bar x=\{x_1,\dots,x_{a+b}\}$ be a set of variables.
Let the matrix elements of ${\sf N}$ have the form
\be{matel-A}
\begin{array}{ll}
{\sf N}_{jk}={\sf N}^{(1)}_j(x_k),& j=1,\dots,a\\
{\sf N}_{jk}={\sf N}^{(2)}_j(x_k),& j=a+1,\dots,a+b,
\end{array}
\ee
where ${\sf N}^{(1)}_j(x)$ and ${\sf N}^{(2)}_j(x)$ are some functions of $x$. Then
\be{Det-A}
\Delta_{a+b}(\bar x)\det_{a+b}{\sf N}=\sum g(\bar x_{\st},\bar x_{\so}) \;
\bigl[\Delta_a(\bar x_{\so})\det_a\bigl({\sf N}^{(1)}_j(x^{\so}_k)\bigr)\bigr]
\cdot \bigl[\Delta_b(\bar x_{\st})\det_b\bigl({\sf N}^{(2)}_j(x^{\st}_k)\bigr)\bigr].
\ee
Here the sum is taken with respect to the partitions $\bar x\Rightarrow\{\bar x_{\so},\bar x_{\st}\}$ with
$\#\bar x_{\so}=a$ and $\#\bar x_{\st}=b$. The notation $x^{\so}_k$ (resp. $x^{\st}_k$) means the
$k$-th element of the subset $x_{\so}$ (resp. $x_{\st}$), and we assume that the elements in every subset
are ordered in the natural order.
\end{prop}

{\sl Proof.} Developing $\det_{a+b}{\sf N}$ with respect to the first $a$ rows we obtain
\be{Det-A-exp}
\det_{a+b}{\sf N}=\sum (-1)^{[P_{\so,\st}]}
\det_a\bigl({\sf N}^{(1)}_j(x^{\so}_k)\bigr)
\cdot \det_b\bigl({\sf N}^{(2)}_j(x^{\st}_k)\bigr),
\ee
where the sum is taken with respect to the partitions $\bar x\Rightarrow\{\bar x_{\so},\bar x_{\st}\}$ with
$\#\bar x_{\so}=a$ and $\#\bar x_{\st}=b$. The symbol
$[P_{\so,\st}]$ means the parity of the permutation $P_{\so,\st}$ mapping the union of the subsets $\{\bar x_{\so},\bar x_{\st}\}$
into the naturally ordered set $\bar x$: $P_{\so,\st}(\{\bar x_{\so},\bar x_{\st}\})=\bar x$. On the other hand the product
$\bigl(\Delta(\bar x)\bigr)^{-1}$ coincides up to a constant
with the Vandermonde determinant of variables $\bar x$. Reordering the rows of this determinant by the permutation
$P^{-1}_{\so,\st}$ we obtain
\be{Perm-delta0}
\Delta_{a+b}(\bar x)= (-1)^{[P_{\so,\st}]}g(\bar x_{\st},\bar x_{\so}) \Delta_a(\bar x_{\so})\Delta_b(\bar x_{\st}),
\ee
and hence,
\be{Perm-delta}
(-1)^{[P_{\so,\st}]}= \frac{g(\bar x_{\st},\bar x_{\so}) \Delta_a(\bar x_{\so})\Delta_b(\bar x_{\st})}
{\Delta_{a+b}(\bar x)}
\ee
for the arbitrary partition $\bar x\Rightarrow\{\bar x_{\so},\bar x_{\st}\}$. Substituting \eqref{Perm-delta} into \eqref{Det-A-exp} we immediately arrive at \eqref{Det-A}.\qed

It is easy to see that \eqref{Part-sum-4} is a particular case of \eqref{Det-A}. Indeed, define a set $\bar x$ as
in \eqref{x-def}
\be{x-def1}
\bar x=\{\blab,\bmuc\}=\{\lab_1,\dots\lab_a,\muc_1,\dots,\muc_b\},
\ee
and let us set in \eqref{matel-A}
\be{matel-A-NN}
{\sf N}^{(1)}_j(x)={\sf N}^{(u)}(\lac_j,x), \qquad
{\sf N}^{(2)}_j(x)={\sf N}^{(v)}(\mub_j,x).
\ee
Then due to proposition~\ref{GenMat}
\begin{multline}\label{Det-A1}
\Delta'_a(\blac)\Delta'_b(\bmub)\Delta_{a+b}(\bar x)\det_{a+b}{\sf N}\\
=\sum g(\bar x_{\st},\bar x_{\so}) \;
\bigl[\Delta'_a(\blac)\Delta_a(\bar x_{\so})\det_a\bigl({\sf N}^{(u)}(\lac_j,x^{\so}_k)\bigr)\bigr]
\cdot \bigl[\Delta'_b(\bmub)\Delta_b(\bar x_{\st})\det_b\bigl({\sf N}^{(v)}(\mub_j,x^{\st}_k)\bigr)\bigr]\\
=\sum g(\bar x_{\st},\bar x_{\so})\hat{\sf L}_{a}(\bar x_{\so}|\blac) \hat{\sf M}_{b}(\bar x_{\st}|\bmub).
\end{multline}
If we set $\bar x_{\so}=\{\blab_{\st},\bmuc_{\so}\}$ and $\bar x_{\st}=\{\blab_{\so},\bmuc_{\st}\}$, then
we reproduce the sum in the r.h.s. \eqref{Part-sum-4}. Thus, we finally obtain the result \eqref{FF1-ans-m1}.

\subsection{The case $\kappa_3/\kappa_1=q^2$}

Consider now the case $\kappa_3/\kappa_1=q^2$. Let us denote the corresponding
scalar product by $S^{(q^2)}_{a,b}$. Substituting \eqref{Gq2} into \eqref{Part-sum-1} we obtain
 \begin{multline}\label{Part-sum-q2}
S^{(q^2)}_{a,b}=\mathfrak{p}(\bmub)\sum (-1)^{n_{\so}}
  f(\bmuc_{\st},\blab_{\st})f(\blab_{\so},\blab_{\st}) f(\bmuc_{\st},\bmuc_{\so})
  t(\bmuc_{\so},\blab_{\so})h(\blab_{\so},\blab_{\so})h(\bmuc_{\so},\bmuc_{\so})
\num
 \times \mathfrak{p}(\bmuc_{\so})\mathfrak{p}(\blab_{\st}) {\sf L}_{a}(\{\blab_{\st},\bmuc_{\so}\}|\blac)
 {\sf M}_{b}(\{\blab_{\so},\bmuc_{\st}\}|\bmub).
\end{multline}
Acting exactly in the same manner as before we obtain
 \begin{multline}\label{Part-sum-q4}
S^{(q^2)}_{a,b}= \mathfrak{p}(\bmub)C_h\sum
g(\bmuc_{\st},\blab_{\st})g(\blab_{\so},\blab_{\st}) g(\bmuc_{\st},\bmuc_{\so})
  g(\blab_{\so},\bmuc_{\so})
\num
 \times \mathfrak{p}(\bmuc_{\so})\mathfrak{p}(\blab_{\st})
 \hat{\sf L}_{a}(\{\blab_{\st},\bmuc_{\so}\}|\blac) \hat{\sf M}_{b}(\{\blab_{\so},\bmuc_{\st}\}|\bmub).
\end{multline}
Comparing the obtained equation with \eqref{Part-sum-4} we see that the difference is very small, namely, we
have the additional factor $\mathfrak{p}(\bmuc_{\so})\mathfrak{p}(\blab_{\st})$ in \eqref{Part-sum-q2}. This
factor can be easily included into determinants. Consider the matrix $\widetilde{\sf N}$ \eqref{mat-tN}.
Then due to proposition~\ref{GenMat} we have
\begin{multline}\label{Det-Aq}
\Delta'_a(\blac)\Delta'_b(\bmub)\Delta_{a+b}(\bar x)\det_{a+b}\widetilde{\sf N}
=\sum g(\bar x_{\st},\bar x_{\so}) \;
\bigl[\Delta'_a(\blac)\Delta_a(\bar x_{\so})\det_a\bigl(x^{\so}_k{\sf N}^{(u)}(\lac_j,x^{\so}_k)\bigr)\bigr]\\
\hspace{30mm}\times \bigl[\Delta'_b(\bmub)\Delta_b(\bar x_{\st})\det_b\bigl({\sf N}^{(v)}(\mub_j,x^{\st}_k)\bigr)\bigr]\\
=\sum g(\bar x_{\st},\bar x_{\so}) \mathfrak{p}(\bar x_{\so})
\bigl[\Delta'_a(\blac)\Delta_a(\bar x_{\so})\det_a\bigl({\sf N}^{(u)}(\lac_j,x^{\so}_k)\bigr)\bigr]
%
\bigl[\Delta'_b(\bmub)\Delta_b(\bar x_{\st})\det_b\bigl({\sf N}^{(v)}(\mub_j,x^{\st}_k)\bigr)\bigr]\\
=\sum g(\bar x_{\st},\bar x_{\so})\mathfrak{p}(\bar x_{\so})\hat{\sf L}_{a}(\bar x_{\so}|\blac) \hat{\sf M}_{b}(\bar x_{\st}|\bmub).
\end{multline}
After specification $\bar x_{\so}=\{\blab_{\st},\bmuc_{\so}\}$ and $\bar x_{\st}=\{\blab_{\so},\bmuc_{\st}\}$
we reproduce the sum in the r.h.s. of \eqref{Part-sum-q2}. Thus, we finally arrive
at \eqref{FF1-ans-mq}.

\section{Discussions\label{S-FinDis}}

We have described the derivation of the determinant formulas for the scalar product of twisted and usual on-shell Bethe vectors in the
generalized $GL(3)$-based model with the trigonometric $R$-matrix. Comparing this derivation with the one given in \cite{BelPRS12b,BelPRS13a} for
the case of the $GL(3)$-invariant $R$-matrix we see that not only the general line, but even the details
are almost the same. The differences are very small and mostly occur in the presence of degrees of the deformation parameter $q$. However,
these  negligible differences eventually lead to drastic consequences. Namely, we succeeded in obtaining the determinant representation for the
form factor of the operator $T_{22}(z)$ only. The other two form factors of the diagonal entries  were unattainable within the framework of our scheme.
Indeed, in order to calculate the form factors of  $T_{11}(z)$ and $T_{33}(z)$ we should be able to compute $\kappa$-derivatives of the
function ${\sf G}^{(\kappa)}_{n_{\so}}$ \eqref{G-kappa1} at $\bar\kappa=1$
\be{G-kappa-der}
   \frac{d{\sf G}^{(\kappa)}_{n_{\so}}}{d\kappa_s}\Bigr|_{\bar\kappa=1}=\pm\sum
 n_{\rm i}\,q^{2n_{\rm i}}f(\blab_{\rm i},\blab_{\rm iv})
 f(\bmuc_{\rm iv},\bmuc_{\rm i})\Izerl_{n_{\rm iv}}(\blab_{\rm iv}|\bmuc_{\rm iv})\;
\Izerr_{n_{\rm i}}(\bmuc_{\rm i}|\blab_{\rm i}q^{-2}),
\ee
where $s=1,3$. In the case of the $GL(3)$-invariant $R$-matrix an analog of this sum over partitions was calculated in \cite{BelPRS13a}. However,
in the $q$-deformed case the sum \eqref{G-kappa-der} is not described by summation lemma~\ref{Wau} or its corollaries. In particular,
we have the $q$-number $[n_{\rm i}]$ in the identity \eqref{Ident-G0i}, while \eqref{G-kappa-der} contains the usual number $n_{\rm i}$.

Curiously, for the other matrix elements $T_{ij}(z)$ we have a similar situation: we cannot obtain determinant representation for the  usual form factors, however it can be done for twisted form factors if the twist parameters are related to the deformation parameter $q$. In particular, we have found a determinant representation for the twisted form factor of the operator $T_{12}(z)$, provided $\kappa_3/\kappa_1=q$ and $\kappa_2$ is arbitrary. For completeness we
describe this result below; however, we do not give the detailed derivation, because its main steps coincide with the ones of \cite{PakRS14d}, while the
peculiarities  arising due to the $q$-deformation are already described in the present paper.

Twisted form factor of $T_{12}(z)$ is defined as
 \be{SP-deFF-gen12}
 \mathcal{F}_{a,b;\bar\kappa}^{(1,2)}(z)\equiv\mathcal{F}_{a,b;\bar\kappa}^{(1,2)}(z|\blac,\bmuc;\blab,\bmub)=
 \mathbb{C}^{a+1,b}(\blac;\bmuc)T_{12}(z)\mathbb{B}^{a,b}(\blab;\bmub),
 \ee
where  $\mathbb{C}^{a+1,b}(\blac;\bmuc)$ is a twisted on-shell
Bethe vector, while $\mathbb{B}^{a,b}(\blab;\bmub)$ is a standard on-shell
Bethe vector. Note that in the distinction of the case considered above the number of elements in the
set $\blac$ is $a+1$.

If $\kappa_3/\kappa_1=q$, then a determinant representation for $\mathcal{F}_{a,b;\bar\kappa}^{(1,2)}(z)$ has the following form:
\be{FF1-ans-12}
\mathcal{F}_{a,b;\bar\kappa}^{(1,2)}(z)= z \mathfrak{p}(\bmub)\mathfrak{p}(\blab) C_h h(\bmuc,z)h(z,\blab)\Delta'_{a+1}(\blac)
 \Delta'_b(\bmub)\Delta_{a+b+1}(\bar x)\;\det_{a+b+1}{\sf N}^{(12)},
\ee
where
\be{mat-N12}
\begin{array}{ll}
{\sf N}^{(12)}_{j,k}={\sf N}^{(u)}(\lac_j,x_k),&j=1,\dots,a+1,\\
{\sf N}^{(12)}_{j+a+1,k}={\sf N}^{(v)}(\mub_j,x_k),&j=1,\dots,b,
\end{array}
\qquad \{x_1,\dots,x_{a+b+1}\}=\{\lab_1,\dots\lab_a,z,\muc_1,\dots,\muc_b\},
\ee
The matrix elements ${\sf N}^{(v)}(\mub_j,x_k)$ are given by \eqref{Nv-def}, and the matrix elements ${\sf N}^{(u)}(\lac_j,x_k)$ are given by
\eqref{Nu-def}, where one should replace $(-1)^{a-1}$ by $(-1)^{a}$. This replacement is related to the fact that now the set
$\blac$ consists of $a+1$ elements.

It is worth mentioning that if we consider twisted form factors where $\mathbb{C}^{a,b}(\blac;\bmuc)$ is a usual on-shell
Bethe vector, while $\mathbb{B}^{a,b}(\blab;\bmub)$ is a twisted on-shell Bethe vector, then in order to obtain a determinant representation  one should take inverse ratios
of the twist parameters. For example, the  determinant representations for these twisted form factors of the operators $T_{11}(z)$
and $T_{33}(z)$ exist, if $\kappa_3/\kappa_1=q^{-2}$. This is not surprising, because one can say that $T_{\bar\kappa}(u)$ is the original monodromy matrix,
and then $T(u)$ becomes a twisted monodromy matrix with the twist parameters $\kappa_i^{-1}$. In other words only the relative twist is important.

This observation makes it impossible to expand over the twisted form factors with the fixed ratios of the twist parameters.
Generically twisted form factors can be used for the expansion of complex operators (see. e.g. \cite{KitMST05}). For instance,
if we need to calculate the matrix element of the product $T_{11}(z_1)T_{33}(z_2)$, then we can expand it into a form factor series as follows:
 \begin{multline}\label{expan}
  \mathbb{C}^{a,b}(\blac;\bmuc)T_{11}(z_1)T_{33}(z_2)\mathbb{B}^{a,b}(\blab;\bmub)\\
  =\sum_{\{\bla,\bmu\}}
   \frac{1}{\|\mathbb{B}^{a,b}(\bla;\bmu)\|^2}\;\mathbb{C}^{a,b}(\blac;\bmuc)T_{11}(z_1)\mathbb{B}^{a,b}(\bla;\bmu)\cdot
   \mathbb{C}^{a,b}(\bla;\bmu)T_{33}(z_2)\mathbb{B}^{a,b}(\blab;\bmub).
 \end{multline}
Here the vectors $\mathbb{B}^{a,b}(\bla;\bmu)$ (and their dual $\mathbb{C}^{a,b}(\bla;\bmu)$) are not necessarily  usual on-shell Bethe vectors.
They might be twisted on-shell Bethe vectors as well.  In the last case we are just dealing with the twisted form factors.
Suppose that $\mathbb{C}^{a,b}(\bla;\bmu)$ in \eqref{expan} is a twisted on-shell dual Bethe vector with $\kappa_3/\kappa_1=q^2$.
Then we know the determinant representation for this twisted form factor. But in this case the twisted on-shell Bethe vector
$\mathbb{B}^{a,b}(\bla;\bmu)$ also has $\kappa_3/\kappa_1=q^2$, and hence, we do not know a determinant representation for this
twisted form factor. Thus, the determinant representations obtained above cannot be used in \eqref{expan}.

In conclusion of this discussion we would like to comment on the zero modes method developed in
\cite{PakRS15a}  for calculating form factors  in $GL(3)$-invariant models. There it was proved that the  form factors of all matrix elements $T_{ij}(z)$ can be obtained from one initial form factor, say  $\mathcal{F}_{a,b}^{(2,2)}(z)$,  by taking the special limits of the Bethe parameters.
Since in the $q$-deformed case the determinant formula for the form factor $\mathcal{F}_{a,b}^{(2,2)}(z)$ is known, one could hope to apply the same approach
in order to derive determinants in the $q$-deformed case for  other operators $T_{ij}(z)$. However, the zero modes method fails for the models with the trigonometric $R$-matrix.
It is possible that there exists some modification of this method, which makes it applicable to such models.  But to date such a modification is not known.

In the light of the foregoing, the existence of the determinant formula for the form factor of $T_{22}(z)$  appears to be
a miraculous exception. At the same time
this fact allows us to hope that determinant formulas for other form factors also exist and can be obtained by certain improvement of the method
described in the present paper.

\section*{Acknowledgements}
It is a great pleasure for me to thank my colleagues S. Pakuliak and E. Ragousy for numerous and fruitful discussions.
This work was supported by the RSF under a grant 14-50-00005.

\appendix

\section{Proof of lemma~\ref{Long-Det}\label{A-Long-Det}}

Consider a partition of a set $\bar\gamma$ into two subsets $\{\bar\gamma_{\so},\bar\gamma_{\st}\}$ with $\#\bar\gamma_{\so}=n$,
$n=0,\dots,m$. Let us write explicitly the Izergin determinant $\Izerlr_m(\{\bar\gamma_{\so}q^{-2}, \bar\gamma_{\st}\}|\bar \xi)$:
\begin{multline}\label{K-shift}
\Izerlr_m(\{\bar\gamma_{\so}q^{-2}, \bar\gamma_{\st}\}|\bar \xi)=q^{-n\mp n}
\mathfrak{p}^{\ell,r}\Delta'_m(\bar\xi)\Delta_n(\bar\gamma_{\so}q^{-2})
\Delta_{m-n}(\bar\gamma_{\st})\;g(\bar\gamma_{\st},\bar\gamma_{\so}q^{-2})\\
\times h(\bar\gamma_{\so}q^{-2},\bar \xi)h(\bar\gamma_{\st},\bar \xi)\det_m\Bigl(t(\gamma^{\so}_kq^{-2},\xi_j)\Bigr|
t(\gamma^{\st}_k,\xi_j)\Bigr).
\end{multline}
Here  the symbols
$\gamma^{\so}_k$ (resp. $\gamma^{\st}_k$) denote the $k$-th element of the subset $\gamma_{\so}$
(resp. $\gamma_{\st}$). Recall that the elements in every subset are ordered in the natural order.  The matrix elements in the first $n$ columns of the determinant in \eqref{K-shift} are equal
to $t(\gamma^{\so}_kq^{-2},\xi_j)$, while in the last $(m-n)$ columns we have $t(\gamma^{\st}_k,\xi_j)$. Due to  relations \eqref{formulas}
the equation \eqref{K-shift} can be written in the form
\begin{equation}\label{K-shift1}
\Izerlr_m(\{\bar\gamma_{\so}q^{-2}, \bar\gamma_{\st}\}|\bar \xi)=q^{\mp n}
\mathfrak{p}^{\ell,r}\frac{\Delta'_m(\bar\xi)\Delta_n(\bar\gamma_{\so})
\Delta_{m-n}(\bar\gamma_{\st})h(\bar\gamma_{\st},\bar \xi)}{h(\bar\gamma_{\st},\bar\gamma_{\so})g(\bar\gamma_{\so},\bar \xi)}\;
\det_m\Bigl(t(\xi_j,\gamma^{\so}_k)\Bigr|
t(\gamma^{\st}_k,\xi_j)\Bigr).
\end{equation}
Hence, we obtain
\begin{multline}\label{K-shift2}
\Izerlr_m(\{\bar\gamma_{\so}q^{-2}, \bar\gamma_{\st}\}|\bar \xi)f(\bar\gamma_{\st},\bar\gamma_{\so})f(\bar \xi,\bar\gamma_{\so})
=\mathfrak{p}^{\ell,r}\Delta'_m(\bar\xi)\Delta_n(\bar\gamma_{\so})
\Delta_{m-n}(\bar\gamma_{\st})g(\bar\gamma_{\st},\bar\gamma_{\so})\\
\times \det_m\Bigl((-1)^mq^{\mp 1} t(\xi_j,\gamma^{\so}_k)h(\bar \xi,\gamma^{\so}_k)\;\Bigr|\;
t(\gamma^{\st}_k,\xi_j)h(\gamma^{\st}_k,\bar \xi)\Bigr).
\end{multline}

Consider now the r.h.s. of \eqref{SumDet1} as a functional of functions $\phi_1(\gamma)$ and $\phi_2(\gamma)$. Clearly this
functional is linear with respect to any $\phi_1(\gamma_k)$ and $\phi_2(\gamma_k)$, $k=1,\dots,m$. Therefore we have
\begin{multline}\label{Det-rhs}
\mathfrak{p}^{\ell,r}\;\Delta'_m(\bar\xi)\Delta_m(\bar\gamma)
\det_m\Bigl(\phi_2(\gamma_k)t(\gamma_k,\xi_j)h(\gamma_k,\bar\xi)+q^{\mp 1}(-1)^m \phi_1(\gamma_k)t(\xi_j,\gamma_k)h(\bar\xi,\gamma_k)\Bigr)\\
=\sum \phi_1(\bar\gamma_{\so})\phi_2(\bar\gamma_{\st}) \;A(\bar\gamma_{\so},\bar\gamma_{\st}),
\end{multline}
where the sum is taken over all the partitions $\bar\gamma\Rightarrow\{\bar\gamma_{\so},\bar\gamma_{\st}\}$, and
the coefficients $A(\bar\gamma_{\so},\bar\gamma_{\st})$ do not depend on $\phi_1$ and $\phi_2$. In order to find
$A(\bar\gamma_{\so},\bar\gamma_{\st})$ corresponding to fixed subsets $\bar\gamma_{\so}$ and $\bar\gamma_{\st}$ we should set
in the l.h.s. of \eqref{Det-rhs}
\be{C-null}
\begin{array}{ll}
\phi_1(\gamma_k)=0,&\text{if}\quad \gamma_k\in\bar\gamma_{\st},\\
\phi_2(\gamma_k)=0,&\text{if}\quad \gamma_k\in\bar\gamma_{\so}.
\end{array}
\ee
Due to the symmetry of the l.h.s. of \eqref{Det-rhs} with respect to all $\gamma_k$ we can move all columns corresponding
to the elements $\gamma_k\in\bar\gamma_{\so}$ to the left. Simultaneously we should make the same reordering in the product
$\Delta_m(\bar\gamma)$. Then we obtain
\begin{multline}\label{LHS}
A(\bar\gamma_{\so},\bar\gamma_{\st})=\mathfrak{p}^{\ell,r}\Delta'_m(\bar\xi)\Delta_n(\bar\gamma_{\so})
\Delta_{m-n}(\bar\gamma_{\st})g(\bar\gamma_{\st},\bar\gamma_{\so})\\
\times \det_m\Bigl((-1)^mq^{\mp 1} t(\xi_j,\gamma^{\so}_k)h(\bar \xi,\gamma^{\so}_k)\;\Bigr|\;
t(\gamma^{\st}_k,\xi_j)h(\gamma^{\st}_k,\bar \xi)\Bigr).
\end{multline}
Comparing this equation with \eqref{K-shift2}  we arrive at the statement of lemma.\qed

\section{Proof of lemma~\ref{Wau}\label{A-Wau}}

The proof is based on  induction over $n$. It is convenient to denote the l.h.s. and the r.h.s. of \eqref{Ident-GG} respectively
by $\Lambda^{(l)}_n(\bar\alpha|\bar\beta)$ and $\Lambda^{(r)}_n(\bar\alpha|\bar\beta)$.
Then we consider the residues of these rational functions in the poles. The last one
occurs in the points $\alpha_i=\beta_j$ and $\alpha_i=\beta_jq^{-2}$, $i,j=1,\dots,n$. The goal is to prove  that the residues in the above poles
can be expressed in terms of $\Lambda^{(l)}_{n-1}(\bar\alpha|\bar\beta)$ and $\Lambda^{(r)}_{n-1}(\bar\alpha|\bar\beta)$. This will
give us the induction step.

First of all observe that at $n_{\so}$ and $n_{\st}$ fixed the sum over partitions of the set $\bar\alpha$ can be calculated explicitly via lemma~\ref{main-ident}
\begin{equation}\label{sum-a}
\sum_{\bar\alpha\Rightarrow\{\bar\alpha_{\so},\;\bar\alpha_{\st}\}}
f(\bar\alpha_{\so},\bar\alpha_{\st})   \Izerr_{n_{\so}}(\bar\beta_{\so}|\bar\alpha_{\so})\Izerl_{n_{\st}}(\bar\alpha_{\st}|\bar\beta_{\st}q^{-2})
  =(-q)^{-n_{\st}} f(\bar\alpha,\bar\beta_{\st}q^{-2})
  \Izerl_{n}(\{\bar\beta_{\st}q^{-4},\bar\beta_{\so}\} |\bar\alpha).
       \end{equation}
Thus, the l.h.s. of  \eqref{Ident-GG} takes the form
\begin{equation}\label{L-l}
\Lambda^{(l)}_n(\bar\alpha|\bar\beta)=\sum_{\bar\beta\Rightarrow\{\bar\beta_{\so},\;\bar\beta_{\st}\}}
  (-1)^{n_{\st}}q^{n_{\so}-n_{\st}}f(\bar\beta_{\so},z)f(\bar\beta_{\st},\bar\beta_{\so}) f(\bar\alpha,\bar\beta_{\st}q^{-2})
  \Izerl_{n}(\{\bar\beta_{\st}q^{-4},\bar\beta_{\so}\} |\bar\alpha).
       \end{equation}
We will consider $\Lambda^{(l)}_n(\bar\alpha|\bar\beta)$ and $\Lambda^{(r)}_n(\bar\alpha|\bar\beta)$ as functions of $\alpha_n$ and other fixed variables. They are rational functions of $\alpha_n$ and they have poles at\footnote{The poles of the Izergin determinant at $\alpha_n=\beta_kq^{-4}$ are compensated by the zeros of the product $f(\bar\alpha,\bar\beta_{\st}q^{-2})$.}
$\alpha_n=\beta_k$ and $\alpha_n=\beta_kq^{-2}$, $k=1,\dots,n$. Consider the properties of these functions in the mentioned poles.
Due to the symmetry of both functions over $\bar\beta$ it is enough to consider the case $\beta_k=\beta_n$.

Let $\alpha_n=\beta_n$. The pole of $\Lambda^{(l)}_n(\bar\alpha|\bar\beta)$ occurs if and only if $\beta_n\in\bar\beta_{\so}$.
Let $\bar\beta_{\so}=\{\beta_n,\bar\beta_{\so'}\}$. Then setting $n_{\so'}=n_{\so}-1$ and using the property of the Izergin determinant
\eqref{K-Res} we obtain
\begin{multline}\label{L-lres1}
\Lambda^{(l)}_n(\bar\alpha|\bar\beta)\Bigr|_{\alpha_n=\beta_n}=q\sum_{\bar\beta\Rightarrow\{\bar\beta_{\so'},\;\bar\beta_{\st}\}}
  (-1)^{n_{\st}}q^{n_{\so'}-n_{\st}}f(\beta_{n},z)f(\bar\beta_{\so'},z) f(\bar\beta_{\st},\beta_{n})
  f(\bar\beta_{\st},\bar\beta_{\so'})\num
  \times  f(\alpha_n,\bar\beta_{\st}q^{-2})f(\bar\alpha_n,\bar\beta_{\st}q^{-2}) f(\beta_n,\alpha_n)
  f(\alpha_n,\bar\alpha_n) f(\bar\beta_{\so'},\beta_{n})f(\bar\beta_{\st}q^{-4},\beta_{n})\num
  \times  \Izerl_{n-1}(\{\bar\beta_{\st}q^{-4},\bar\beta_{\so'}\} |\bar\alpha_n)+ {\rm reg}.
       \end{multline}
The terms $f(\beta_{n},z)$  and $f(\beta_n,\alpha_n)f(\alpha_n,\bar\alpha_n)$  can be moved out off the sum. The  terms $f(\bar\beta_{\st},\beta_{n})$ and $f(\bar\beta_{\so'},\beta_{n})$ combine into $f(\bar\beta_{n},\beta_{n})$ and also can be moved out off the sum. Finally, the  terms
$f(\alpha_n,\bar\beta_{\st}q^{-2})$ and $f(\bar\beta_{\st}q^{-4},\beta_{n})$     cancel each other.  We obtain
\begin{multline}\label{L-lres1a}
\Lambda^{(l)}_n(\bar\alpha|\bar\beta)\Bigr|_{\alpha_n=\beta_n}=q f(\beta_n,\alpha_n)f(\beta_{n},z)f(\bar\beta_{n},\beta_{n})
  f(\alpha_n,\bar\alpha_n)\num
  \times \sum_{\bar\beta\Rightarrow\{\bar\beta_{\so'},\;\bar\beta_{\st}\}}
  (-1)^{n_{\st}}q^{n_{\so'}-n_{\st}}f(\bar\beta_{\so'},z)
  f(\bar\beta_{\st},\bar\beta_{\so'}) f(\bar\alpha_n,\bar\beta_{\st}q^{-2})
  \Izerr_{n-1}(\{\bar\beta_{\st}q^{-4},\bar\beta_{\so'}\} |\bar\alpha_n)+ {\rm reg}.
       \end{multline}
The remaining sum evidently gives $\Lambda^{(l)}_{n-1}(\bar\alpha_n|\bar\beta_n)$, and we arrive at
\begin{equation}\label{L-lres1b}
\Lambda^{(l)}_n(\bar\alpha|\bar\beta)\Bigr|_{\alpha_n=\beta_n}=q f(\beta_n,\alpha_n)f(\beta_{n},z)f(\bar\beta_{n},\beta_{n})
  f(\alpha_n,\bar\alpha_n)\Lambda^{(l)}_{n-1}(\bar\alpha_n|\bar\beta_n)+ {\rm reg}.
       \end{equation}

Let now $\alpha_n=\beta_nq^{-2}$. The pole of $\Lambda^{(l)}_n(\bar\alpha|\bar\beta)$ occurs if and only if $\beta_n\in\bar\beta_{\st}$.
Let $\bar\beta_{\st}=\{\beta_n,\bar\beta_{\st'}\}$. Then setting $n_{\st'}=n_{\st}-1$ and using the property of the Izergin determinant
\eqref{K-red}
 \begin{equation}\label{K-red2}
\Izerl_{n+1}(\{\bar x, q^{-2}z\}|\{\bar y,z\}) =-q\Izerl_{n}(\bar x|\bar y),
\end{equation}
we obtain
\begin{multline}\label{L-lres2}
\Lambda^{(l)}_n(\bar\alpha|\bar\beta)\Bigr|_{\alpha_n=\beta_nq^{-2}}=\sum_{\bar\beta\Rightarrow\{\bar\beta_{\so},\;\bar\beta_{\st'}\}}
  (-1)^{n_{\st'}}q^{n_{\so}-n_{\st'}}f(\bar\beta_{\so},z) f(\beta_{n},\bar\beta_{\so})
  f(\bar\beta_{\st'},\bar\beta_{\so})\num
  \times  f(\alpha_n,\beta_nq^{-2})
  f(\bar\alpha_n,\alpha_n)f(\beta_{n},\bar\beta_{\st'})f(\bar\alpha_n,\bar\beta_{\st'}q^{-2})
  \Izerl_{n-1}(\{\bar\beta_{\st'}q^{-4},\bar\beta_{\so}\} |\bar\alpha_n)+ {\rm reg}.
       \end{multline}
The terms $f(\alpha_n,\beta_nq^{-2})  f(\bar\alpha_n,\alpha_n)$ can be moved out off the sum.  The  terms
 $ f(\beta_{n},\bar\beta_{\so})$ and $f(\beta_{n},\bar\beta_{\st'})$ combine into $f(\beta_{n},\bar\beta_{n})$ and also can be moved out off the sum. The remaining sum evidently gives $\Lambda^{(l)}_{n-1}(\bar\alpha_n|\bar\beta_n)$, and we arrive at
\begin{equation}\label{L-lres2a}
\Lambda^{(l)}_n(\bar\alpha|\bar\beta)\Bigr|_{\alpha_n=\beta_nq^{-2}}=
f(\alpha_n,\beta_nq^{-2})  f(\bar\alpha_n,\alpha_n) f(\beta_{n},\bar\beta_{n})\Lambda^{(l)}_{n-1}(\bar\alpha_n|\bar\beta_n)
+ {\rm reg}.
       \end{equation}

Consider now the properties of $\Lambda^{(r)}_n(\bar\alpha|\bar\beta)$ in the same points. Obviously
\begin{multline}\label{L-r-repr}
\Lambda^{(r)}_n(\bar\alpha|\bar\beta)
  =-q\alpha_n\beta_n t(\alpha_n,\beta_n)h(\alpha_n,z)g(\beta_n,z)t(\alpha_n,\bar\beta_n)t(\bar\alpha_n,\beta_n)\\
  h(\alpha_n,\bar\alpha_n)h(\bar\alpha_n,\alpha_n)h(\beta_n,\bar\beta_n)h(\bar\beta_n,\beta_n)
  \;\Lambda^{(r)}_{n-1}(\bar\alpha_n|\bar\beta_n),
       \end{multline}
where we have used $h(x,x)=x$. Taking the limit $\alpha_n\to\beta_n$ we obtain
\begin{multline}\label{L-rRes2}
\Lambda^{(r)}_n(\bar\alpha|\bar\beta)\Bigr|_{\alpha_n=\beta_n}
  =-q\alpha_n^2 \frac{g(\alpha_n,\beta_n)}{h(\alpha_n,\alpha_n)}
  h(\beta_n,z)g(\beta_n,z)\frac{g(\beta_n,\bar\beta_n)}{h(\beta_n,\bar\beta_n)}
  \frac{g(\bar\alpha_n,\alpha_n)}{h(\bar\alpha_n,\alpha_n)}\num
 \times h(\alpha_n,\bar\alpha_n)h(\bar\alpha_n,\alpha_n)h(\beta_n,\bar\beta_n)h(\bar\beta_n,\beta_n)
  \;\Lambda^{(r)}_{n-1}(\bar\alpha_n|\bar\beta_n)+\rm{reg}\num
  =q f(\beta_n,\alpha_n)f(\beta_{n},z)f(\bar\beta_{n},\beta_{n})
  f(\alpha_n,\bar\alpha_n)\Lambda^{(r)}_{n-1}(\bar\alpha_n|\bar\beta_n)+ {\rm reg}.
       \end{multline}

Similarly, setting  $\alpha_n\to\beta_nq^{-2}$ we obtain
\begin{multline}\label{L-rRes3}
\Lambda^{(r)}_n(\bar\alpha|\bar\beta)\Bigr|_{\alpha_n=\beta_nq^{-2}}
  =-q^{-1}\beta_n^2 \frac{g(\beta_nq^{-2},\beta_n)}{h(\alpha_n,\beta_n)}
  h(\beta_nq^{-2},z)g(\beta_n,z)t(\beta_nq^{-2},\bar\beta_n)
  t(\bar\alpha_n,\alpha_nq^2)\num
 \times h(\alpha_n,\bar\alpha_n)h(\bar\alpha_n,\alpha_n)h(\beta_n,\bar\beta_n)h(\bar\beta_n,\beta_n)
  \;\Lambda^{(r)}_{n-1}(\bar\alpha_n|\bar\beta_n)+\rm{reg}.
  \end{multline}
Using \eqref{formulas} and
\be{f}
\frac{q^{-1}\beta_n}{h(\alpha_n,\beta_n)}=f(\alpha_n,\beta_nq^{-2}) \Bigr|_{\alpha_n=\beta_nq^{-2}},
\ee
we immediately recast \eqref{L-rRes3} as follows
  \begin{equation}\label{L-rRes4}
  \Lambda^{(r)}_n(\bar\alpha|\bar\beta)\Bigr|_{\alpha_n=\beta_nq^{-2}}
  =f(\alpha_n,\beta_nq^{-2})  f(\bar\alpha_n,\alpha_n) f(\beta_{n},\bar\beta_{n})
  \Lambda^{(r)}_{n-1}(\bar\alpha_n|\bar\beta_n)+ {\rm reg}.
       \end{equation}

Now everything is ready for induction over $n$.
Assume that
\be{nn1}
  \Lambda^{(l)}_{n-1}(\bar\alpha_n|\bar\beta_n)=  \Lambda^{(r)}_{n-1}(\bar\alpha_n|\bar\beta_n).
  \ee
It is straightforward to check that \eqref{nn1} does hold for $n=2$. Due to the induction assumption and due to equations \eqref{L-lres1b}, \eqref{L-lres2a}, \eqref{L-rRes2}, and
\eqref{L-rRes4} the difference $\Lambda^{(l)}_{n}(\bar\alpha|\bar\beta)-\Lambda^{(r)}_{n}(\bar\alpha|\bar\beta)$ is a bounded function of $\alpha_n$ in the whole complex plain. It is also easy to see that this function is bounded as $\alpha_n\to\infty$. Hence,
it is a constant with respect to $\alpha_n$. But this constant vanishes at $\alpha_n=0$, hence,
$\Lambda^{(l)}_{n}(\bar\alpha|\bar\beta)-\Lambda^{(r)}_{n}(\bar\alpha|\bar\beta)=0$.\qed

\end{document}